 \documentclass[aps,pra,notitlepage,floatfix,twocolumn]{revtex4-1}
\usepackage{amssymb, graphicx,subfigure}
\usepackage{enumerate,tensor}
\usepackage{amsmath,bm}
\usepackage{dcolumn}
\usepackage{color}
\usepackage{natbib} 
\usepackage[latin1]{inputenc}
\usepackage{subfigure}
\usepackage{textcomp}
\usepackage{hyperref}
\usepackage{booktabs}
\usepackage{makecell}
\usepackage{psfrag}
\usepackage{epsfig}

\usepackage[per-mode=symbol]{siunitx}   
\usepackage[version=3]{mhchem}

\DeclareSIUnit\intensity{\watt\per\centi\meter\squared}
\DeclareSIUnit\fieldstrength{\volt\per\centi\meter}
\DeclareSIUnit\kfieldstrength{k\volt\per\centi\meter}

\newcommand{\costheta}{\ensuremath{\left\langle\cos\theta\right\rangle}}%
\newcommand{\melement}[3]{\ensuremath{\left\langle #1 \left|#2\right|#3\right\rangle}}%
\newcommand{\eg}{e.\,g.}%
\newcommand{\Estat}{\ensuremath{\textbf{E}_{\textup{s}}}}%
\newcommand{\Estatabs}{\ensuremath{\text{E}_{\textup{s}}}}%
\newcommand{\ie}{i.\,e.}%
\newcommand{\Ialign}{\textup{I}_{0}}%
\newcommand{\pstate}[4]{\ensuremath{\left|#1_{#2,#3}#4\right>}}%
\newcommand{\ppstate}[4]{\ensuremath{\left|#1_{#2,#3}#4\right>_\text{p}}}%

\makeatletter
\def\paragraph{\@startsection{paragraph}{4}{10pt}{-1.25ex plus -1ex minus -.1ex}{0ex plus 0ex}{\normalsize\textit}}
\renewcommand\@biblabel[1]{#1}
\renewcommand\@makefntext[1]%
{\noindent\makebox[0pt][r]{\@thefnmark\,}#1}
\DeclareRobustCommand\onlinecite{\@onlinecite}
\def\@onlinecite#1{\begingroup\let\@cite\NAT@citenum\citealp{#1}\endgroup}
\def\tagform@#1{\maketag@@@{\ignorespaces#1\unskip\@@italiccorr}}
\let\orgtheequation\theequation
\def\theequation{(\orgtheequation)}
\makeatother

\newcommand{\ket}[1]{\left|#1\right\rangle}

\begin{document}

\title{Rotational dynamics of an asymmetric top molecule in parallel electric and non-resonant  
laser fields}

\author{Juan J.\ Omiste}
\email{omiste@ugr.es}%
\author{Rosario Gonz\'alez-F\'erez}
\email{rogonzal@ugr.es}%

\affiliation{Instituto Carlos I de F\'{\i}sica Te\'orica y Computacional,
and Departamento de F\'{\i}sica At\'omica, Molecular y Nuclear,
  Universidad de Granada, 18071 Granada, Spain} 

\date{\today}
\begin{abstract} 
We present a theoretical study of the rotational dynamics of asymmetry top molecules 
in an electric field and a parallel non-resonant  linearly polarized laser pulse. 
The time-dependent Schr\"odinger equation is solved within the rigid rotor approximation.
Using the benzonitrile molecule as prototype,  we investigate the field-dressed dynamics 
for experimentally accessible field configurations and compare these results to the adiabatic 
predictions. 
We show that for an asymmetric top molecule in parallel fields, the formation of the 
pendular doublets and the  avoided crossings between neighboring levels are the two main 
sources of non-adiabatic effects. We also  provide the field parameters  under which the adiabatic 
dynamics would be achieved. 
\end{abstract}
\maketitle
\section{Introduction}
\label{sec:introduction}

The availability of oriented molecules
provides a wealth of intriguing applications in a variety of molecular sciences, 
such as in chemical reaction dynamics~\cite{brooks:science,brooks:jcp45,loesch:9016,aoiz:chem_phys_lett_289,aquilanti:pccp_7}, 
photoelectron angular distributions~\cite{Bisgaard:Science323:1464,Holmegaard:natphys6,Hansen:PRL106:073001},
 or high-order harmonic generation~\cite{frumker2012,kraus2012,buckbaum}. 
An oriented molecule is characterized by the confinement of the molecular fixed
axes along the laboratory fixed axes and by its dipole
moment pointing in a particular direction. 
Many experimental efforts have been undertaken to control the rotational degree
of freedom, and in particular to orient polar 
molecules~\cite{brooks:science,stolte,loesch:jcp93,friedrich:nature353,friedrich:jpc95,block:prl68,friedrich:prl69,slenczka:prl72,friedrich:jcp111,friedrich:jpca103-a,kupper:prl102,nevo:pccp11}.

Here, we focus 
on a theoretical study of 
the mixed-field orientation technique, which is based on the combination of 
weak dc and strong non-resonant radiative fields~\cite{friedrich:jcp111,friedrich:jpca103-a}.
Strongly oriented/antioriented states could be created by coupling the 
nearly degenerate pair of states with opposite parity forming a
pendular doublet by means of a weak dc field. 
This theoretical prediction is based on an 
adiabatic picture in which 
the turn-on time of the
laser pulse is larger than the molecular rotational period~\cite{seideman2006}.
For asymmetric top molecules, 
a theoretical study  based on a time-independent model  
of their mixed-field orientation pointed out that a fully adiabatic description of this 
process does not reproduce the experimental results~\cite{omiste:pccp2011}.  
We have recently found that a time-dependent description of the
mixed-field orientation of linear molecules is required to explain the experimental
observations~\cite{nielsen:prl2012,omiste:pra2012}. 
Two main sources of non-adiabatic effects were identified for linear molecules:
i) the coupling of the levels forming quasidegenerate
pendular doublets as the laser intensity is increased gives rise to a
transfer of population between them; 
ii) the strongly coupled states from the same
$J$ manifold for tilted fields are driven apart
as the laser intensity is increased in the weak-field regime, provoking   a population
redistribution among them.
In addition, for highly excited states, avoided crossing in the field-dressed spectrum could affect
the rotational dynamics. 
A similar time-dependent study for polar asymmetric top 
molecules is desirable for a correct interpretation of the numerous mixed-field experiments with these
systems~\cite{kupper:prl102,kupper:jcp131}.
Compared to linear molecules, asymmetric tops possess a more dense level structure 
and, when the fields are turned on these levels could be strongly coupled. 
Thus, a more complex rotational dynamics should be expected
for asymmetric tops in combined dc and ac fields.
Let us also mention that several theoretical and experimental studies have investigated 
the relevance of non-adiabatic phenomena on the manipulation of  molecules
with external fields~\cite{bulthuis:jpca101,escribano:pra62,schwettman:jcp123,wohlfart:phys_rev_a_78,kirste:phys_rev_a_79,Meek2009,wall:phys_rev_a_81}.

In this work, we perform a theoretical investigation
of the rotational dynamics of  an asymmetric top in parallel electric
and non-resonant radiative fields within the rigid rotor description.
The time-dependent Schr\"odinger equation is 
solved using experimental field configurations, \ie, first a weak electric field that is switched on
at a constant speed and then a parallel linearly polarized Gaussian 
laser pulse is turned on. 
For several rotational states, we show that under ns-pulses the dynamics is not adiabatic and
it is characterized by the formation 
of the pendular doublets and the numerous avoided crossings of the field-dressed spectrum.
We have proven that due the different time scales associated to each phenomenon, it
might  become experimentally harder to reach the adiabatic limit. 
Increasing the electric field strength helps for the lowest-lying state in a certain irreducible 
representation; whereas for an excited one a proper combination of dc field with the temporal width
 of the pulse is needed to optimize the mixed-field orientation process.

The paper is organized as follows: In \autoref{sec:hamiltonian} we describe
the Hamiltonian of the system, its symmetries and the numerical method used to solve
the time-independent Schr\"odinger equation. 
The main properties of our prototype molecule benzonitrile are provided in \autoref{sec:results}. 
In \autoref{se:ground_states} 
we investigate the mixed-field dynamics of the ground state  of several
irreducible representations, analogously for several excited  states
in \autoref{se:excited_states}. 
%as the laser intensity is increased, and identify the sources of non-adiabatic effects.
We explore the final orientation of these states as the field parameters are varied,
identify the sources of non-adiabatic effects and seek for
the adiabatic regime.
The conclusions are given in \autoref{sec:conclusions}.

\section{The Hamiltonian of an asymmetric top molecule in parallel fields}
\label{sec:hamiltonian}
We consider a polar asymmetric top molecule in  parallel homogeneous
static electric field and non-resonant linearly
polarized laser pulse.  
The polarization of the laser lies along the $Z$-axis of the
laboratory fixed frame (LFF) $(X,Y,Z)$ and the electric field is also parallel to this $Z$-axis. 
We only consider molecules with diagonal polarizability tensors, and the dipole moment parallel to 
the $z$-axis of the molecular fixed frame (MFF) $(x,y,z)$, and 
the smallest moment of inertia is parallel to the $x$-axis.
The LFF and MFF are related by the Euler angles $(\theta, \phi,\chi)$~\cite{zare}.
Within the rigid rotor approximation, the Hamiltonian reads
\begin{equation}
  \label{eq:hamiltonian_total}
  H(t)=H_R+H_\textup{s}(t)+H_L(t),
\end{equation}
where $H_R$ stands for the field-free Hamiltonian
\begin{equation}
H_R=B_xJ_x^2+B_yJ_y^2+B_zJ_z^2 
  \label{eq:hamiltonian_rot}
\end{equation}
with $J_k$ being the projection of the total angular momentum operator $\mathbf{J}$
along the $k$-axis of the MFF with $k=x,y$ and $z$. 
The rotational constant along the MFF $k$-axis is 
$B_k=\cfrac{\hbar^2}{2I_{kk}}$ with 
$I_{kk}$  the moment of inertia with respect to this axis $k$.
We are using the left-handed convention $I^l$~\cite{king_jcp11}, and  
the rotational constant of the considered molecule satisfy $B_z>B_y>B_x$.

The interaction of the electric field $\Estat(t)=\Estatabs(t) \hat{Z}$ with the permanent dipole 
moment, $\boldsymbol{\mu}=\mu\hat{z}$,  reads
\begin{equation}
  \label{eq:hamiltonian_stark}
  H_\textup{s}(t)=-\boldsymbol{\mu}\cdot \Estat(t)=-\mu \Estatabs(t)\cos\theta, 
\end{equation}
where $\Estatabs(t)$ initially depends linearly on time, and once the maximum strength
$ \Estatabs$ is reached, it  is kept constant. The  turning on speed is chosen  so 
that this process is adiabatic.  
Here, we work in the weak  or moderate dc-field regime. 
Thus,  we can neglect the coupling of this field with the molecular polarizability 
and higher order terms.

The interaction of the non-resonant laser field and the molecule 
can be written as~\cite{seideman2006}
\begin{equation}
  \label{eq:hamiltonian_laser}
 H_L(t)=-\cfrac{\textup{I}(t)}{2\epsilon_0c}
 \left(\alpha^{zx}\cos^2\theta+\alpha^{yx}\sin^2\theta\sin^2\chi\right),
\end{equation}
where $\alpha^{km}=\alpha_{kk}-\alpha_{mm}$ are the polarizability
anisotropies, being $\alpha_{kk}$ the polarizability along the molecular axis $k=x,y$ and $z$.
$\epsilon_0$ is the dielectric constant and $c$ the speed of light.
The intensity of the non-resonant laser pulse is $\textup{I}(t)$. 
We analyze Gaussian pulses with intensity
 $\textup{I}(t) = \Ialign\exp\left(-\cfrac{t^2}{2\sigma^2}\right)$, 
$\Ialign$ is  the peak intensity, which is reached at $t=0$, 
and $\sigma$ is related to  full width half maximum (FWHM)  as
$\tau=2\sqrt{2\ln 2}\sigma$.

Based on current mixed-field orientation experiments, we assume a field-free molecule and
turn on the electric field first. Once the maximum dc field strength is reached,
the Gaussian pulse is switched on. 
Since the turning on of the dc field is adiabatic, here we investigate the non-adiabatic 
effects appearing in this second stage.

In this parallel field configuration, the symmetries of the rigid rotor 
Hamiltonian~\eqref{eq:hamiltonian_total}
are the identity, $E$, the two fold rotation around the MFF $z$-axis, arbitrary rotations around the
LFF $Z$ $C_Z(\delta)$, and the reflection on  any plane containing the fields $\sigma_Z$.  
Then, the projection of $\mathbf{J}$ on the $Z$-axis $M$
and the parity of its projection on the $z$-axis, \ie, the parity of $K$,  
are good quantum numbers.
For  $M\ne 0$, there are four irreducible representations and 
the symmetry of reflection on  any plane containing the fields 
implies the well known degeneracy in $|M|$. 
For $M=0$, the wave function can have even and odd parity under these reflections, giving rise to 
two irreducible representations for each parity of $K$. 

\begin{table}
\centering
  \begin{ruledtabular}
\begin{tabular}{lll|c}
%{ccc|c}
 \multicolumn{3}{c}{Parity}& \\
%\cline{1-3}\cline{4-4}
    $C_Z(\delta)$ & $\sigma_{Z}$ & $K$ & Functions\\
\hline
$e$ & $e$ &  $e$, $K=0$ & $|J00\rangle$\\
\hline
% &  & $K$ even\\
$e$ & $e$ & $e$, $K\ne0$ & $\frac{1}{\sqrt{2}}\left(|JK0\rangle +
  (-1)^{K}|J-K0\rangle\right) $\\ %,\quad K\ne 0$\\
\hline
% &  & $K$ even\\
$e$ & $o$ & $e$ & $\frac{1}{\sqrt{2}}\left(|JK0\rangle + (-1)^{K+1}|J-K0\rangle\right)$\\
\hline
% &  & $K$ odd\\
$o$ & $e$ & $o$& $\frac{1}{\sqrt{2}}\left(|JK0\rangle + (-1)^{K}|J-K0\rangle\right)$\\
\hline
% &  & $K$ odd\\
$o$ & $o$ & $o$& $\frac{1}{\sqrt{2}}\left(|JK0\rangle + (-1)^{K+1}|J-K0\rangle\right)$\\
\end{tabular}
  \end{ruledtabular}
\caption{For the states with $M=0$ in parallel fields, functions used in the 
basis set expansion of their wave function. 
% $\sigma_Z$ denotes
%the reflection on any  plane containing the LFF $Z$-axis.
\label{tb:fucntions} }
\end{table}
To solve the time-dependent Schr\"odinger equation of the 
Hamiltonian \eqref{eq:hamiltonian_total},  we employ 
the short iterative Lanczos algorithm for the time propagation~\cite{Beck:phys_rep_324}, and 
a basis set representation for the angular coordinates.
The time step used in short iterative Lanczos algorithm varies from $\delta t=3.5$~fs
to $\delta t=150$~fs for $\tau=0.5$~ns and $\tau=20$~ns, respectively.
In our calculations, the number of vectors in the Krylov space is adapted during the time propagation to
keep the error bellow $10^{-9}$.
For each irreducible representation, we construct a basis using linear combinations of the field-free 
symmetric top eigenfunctions $\ket{JKM}$~\cite{zare} that respect 
the symmetries~\cite{omiste:jcp2011}. 
For the states with $M=0$, we provide in \autoref{tb:fucntions} the basis functions used for each
irreducible representation. 
For the states with $M\ne0$, the basis is formed by the functions $\ket{JKM}$
with the same parity of $K$. 

For reasons of addressability, we label the field-dressed wave function 
using  the field-free notation 
\pstate{J}{K_a}{K_c}{M}$_t$
where $ K_a$ and $K_c$ are the values of $K$ on the limiting 
symmetric top rotor prolate and oblate cases~\cite{king_jcp11}, respectively.
Let us remark that we are using 
the left-handed convention  $I^l$ with $a=z$, $b=y$ and $c=x$.
We have made explicit the dependence on time $t$ of the wave function, but 
not on the field parameters $\Ialign$, $\tau$ and  $\Estatabs$.

To have a better physical insight on the non-adiabatic effects of  the field-dressed 
dynamics, the time-dependent wave function \pstate{J}{K_a}{K_c}{M}$_t$
is expanded in the basis formed by the adiabatic basis at time $t$  
\begin{equation}
  \label{eq:projection_wf}
 \pstate{J}{K_a}{K_c}{M}_t =\sum_{j=0}^N C_{\gamma_j}(t)|\gamma_j\rangle_p
\end{equation}
with $C_{\gamma_j}(t)= {}_p\langle\gamma_j \pstate{J}{K_a}{K_c}{M}_t $, and
$|\gamma_j\rangle_p$ denotes the adiabatic states of Hamiltonian
\eqref{eq:hamiltonian_total} taking the electric field strength and the laser intensities constant.
Note that for each time step $t$,  the time-independent Schr\"odinger equation is solved and 
an adiabatic basis is constructed.
This pendular basis can  be used  to characterize as adiabatic or diabatic the time evolution of a
wave function. Thus, the rotational dynamics  could be considered as fully adiabatic if 
the criterion
\begin{equation*}
  \label{eq:adiabatic_criteria}
  \eta=\cfrac{\hbar\left| 
\tensor[_{\textup{p}}]{\left\langle \gamma_i \left|\cfrac{\partial H_\textup{L}(t)}{\partial t}\right| \gamma_j\right\rangle}{_{\textup{p}}}
\right|}{\left|E_i-E_j\right|^2} \ll 1
\end{equation*}
is satisfied~\cite{ballentine:quantum_mechanics}.

\section{The system}
\label{sec:results}

In this work, we use the benzonitrile molecule (BN) as prototype to illustrate our results. 
Its rotational
constants are $B_{x}=1214$~MHz, $B_{y}=1547$~MHz and $B_{z}=5655$~MHz, the permanent 
dipole moment $\mu=4.515$~D  and the polarizabilities are $\alpha_{xx}=7.49$~\AA$^3$, 
$\alpha_{yy}=13.01$~\AA$^3$ and 
$\alpha_{zz}=18.64$~\AA$^3$~\cite{wohlfart:jms247,PhysRevA.83.023406} 
For the sake of simplicity, we  restrict this study to 
several rotational states of even parity with respect to the 
reflections on $XZ$-plane and under a $\pi$-rotation around the MFF $z$-axis.  
We stress that the observed physical phenomena also
appear for levels within other irreducible representations. 

We consider experimentally accessible field configurations: a linearly polarized 
Gaussian pulse with the FWHM in the nanosecond range and peak intensities 
$ \SI{e11}{\intensity}\le\Ialign\le \SI{e12}{\intensity}$;  and a weak dc field similar to the one
present in a velocity-mapping image spectrometer of few hundreds V/cm. 
Here, we assume that the dc field is switched on slowly
enough to ensure an adiabatic dynamics in this first stage. 
In the experiment, this assumption strongly depends on the  
velocity of the molecules when they reach the velocity-mapping image spectrometer. 
If this tuning-on process is not adiabatic, before the pulse is switched on  
the wave function would be a superposition of several eigenstates of the 
Hamiltonian  \eqref{eq:hamiltonian_total} with 
$\Ialign=\SI{0}{\intensity}$ and the maximum dc field strength
$\Estatabs$. 
As a consequence, the field-dressed dynamics could be more complex since these  states could
belong to different pendular doublets at strong laser intensities.

In a mixed-field orientation experiment, the  measurements are done once the laser pulse
has reached the peak intensity, \ie, at $t=0$ in our theoretical model. 
We first analyze the rotational dynamics as the laser intensity $\textup{I}(t)$ is 
increased. This allows us to understand the different physical phenomena giving rise to a 
non-adiabatic dynamics. 
We also
investigate the orientation at  the peak intensity, \ie, at $t=0$.

 \section{Field-dressed dynamics of the ground state of several irreducible representations}
\label{se:ground_states}

We start analyzing the mixed-field orientation of the  ground states of  the irreducible 
representations with $M=0$  and $3$, 
\ie, the levels $\pstate{0}{0}{0}{0}$ and  $\pstate{3}{0}{3}{3}$. 
We have chosen these states because in mixed-field orientation experiment of BN
they have a 
significant population in the quantum-state selected beam~\cite{kupper:prl102,kupper:jcp131}, and
the knowledge of their field-dressed dynamics is important for these experiments.

\subsection{Dynamics of the state  \pstate{0}{0}{0}{0}$_t$ }
      \begin{figure}[t]
  \centering
  \includegraphics[width=.4\textwidth]{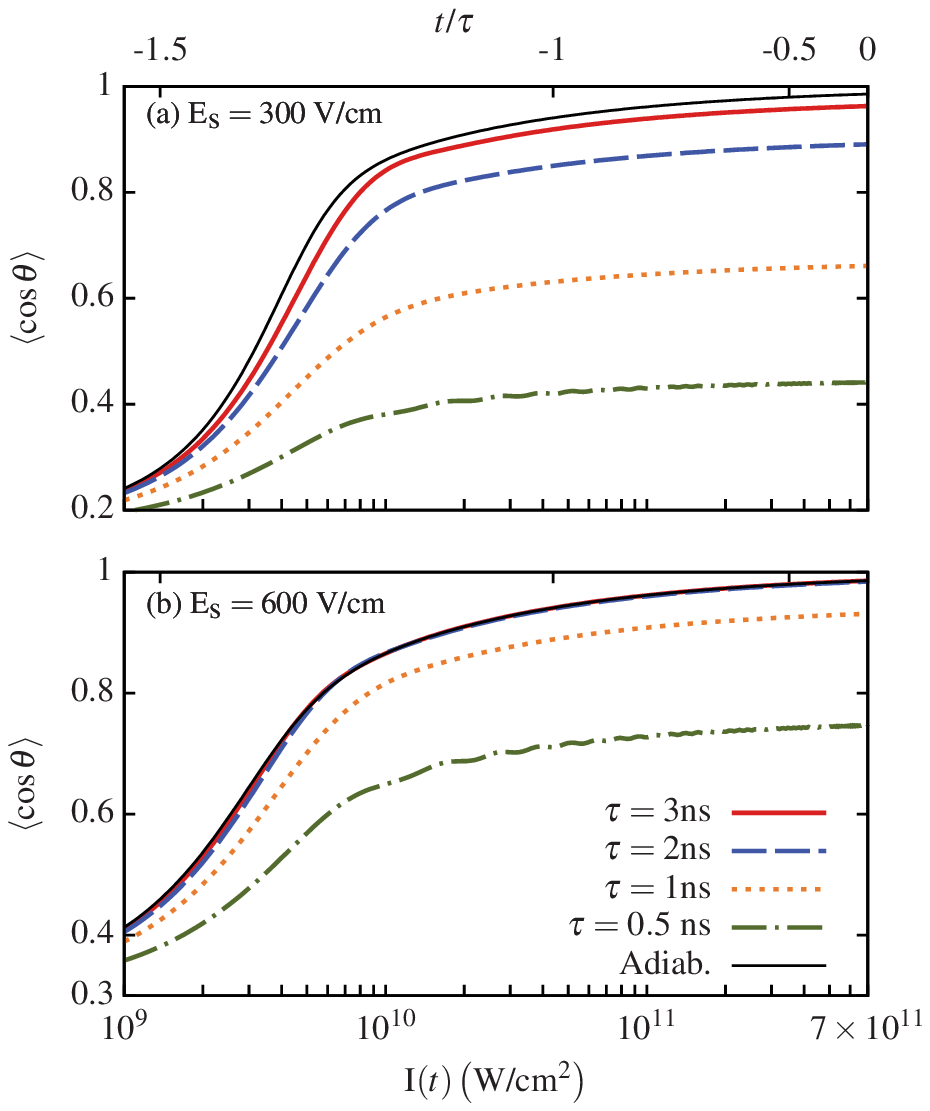}
  \caption{(Color online) For the absolute ground state \pstate{0}{0}{0}{0}$_t$, 
we present the expectation value $\left\langle\cos\theta\right\rangle$ 
  versus the laser intensity $\textup{I}(t)$ for
(a)    $\Estatabs=\SI{300}{\fieldstrength}$  and (b) $\Estatabs= \SI{600}{\fieldstrength}$. 
The  peak intensity is $\Ialign=\SI{7e11}{\intensity}$ and the FWHM are
$\tau=3$~ns (red solid), $\tau=2$~ns (blue long dashed), $\tau=1$~ns (orange dotted) and 
$\tau=0.5$~ns (green
  dashed-dotted). }
  \label{fig:function_fwhm_00_00_00_00}
\end{figure}

\begin{figure}[h]
  \centering
  \includegraphics[width=.32\textwidth]{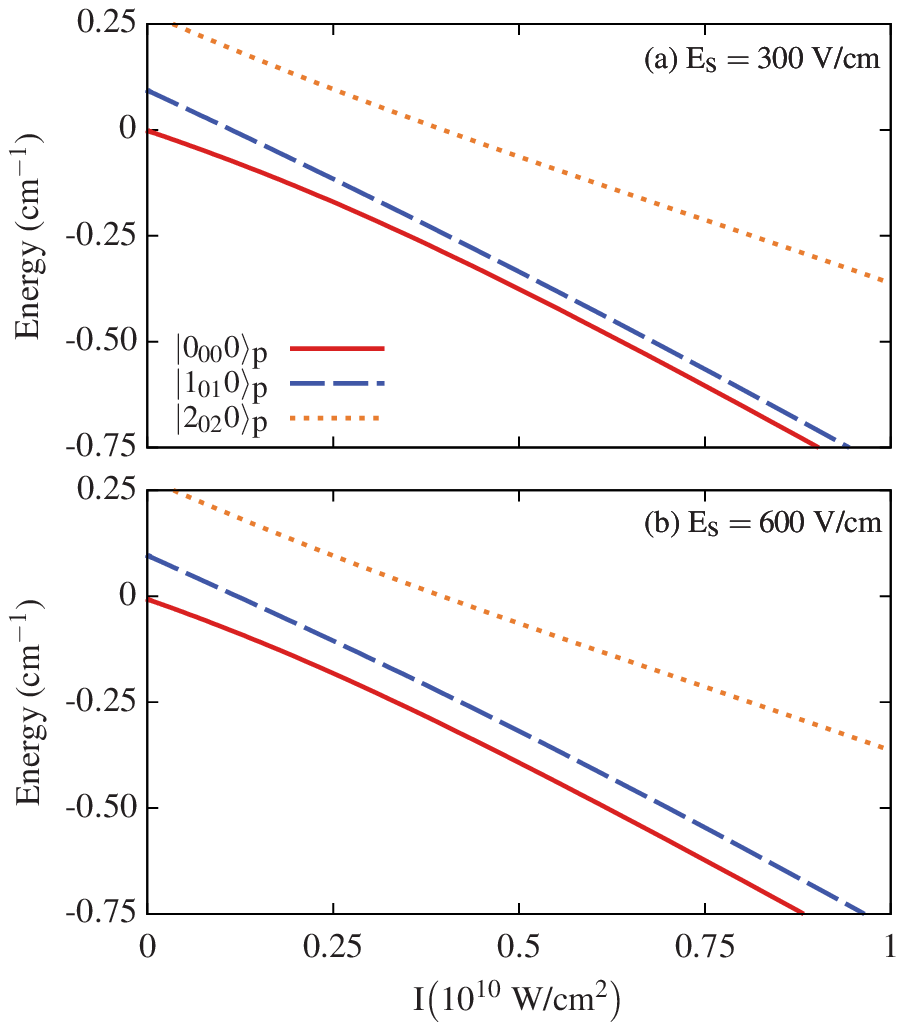}
  \caption{(Color online) 
We present the formation of the  pendular doublet between the states
\ppstate{0}{0}{0}{0} and \pstate{1}{0}{1}{0} in the adiabatic spectrum
for dc field strengths (a) $\Estatabs=\SI{300}{\fieldstrength}$ and (b) $\Estatabs= \SI{600}{\fieldstrength}$.}
  \label{fig:detail_doublet_00_00_00_00}
\end{figure}

\begin{figure}[t]
   \centering
  \includegraphics[width=.4\textwidth]{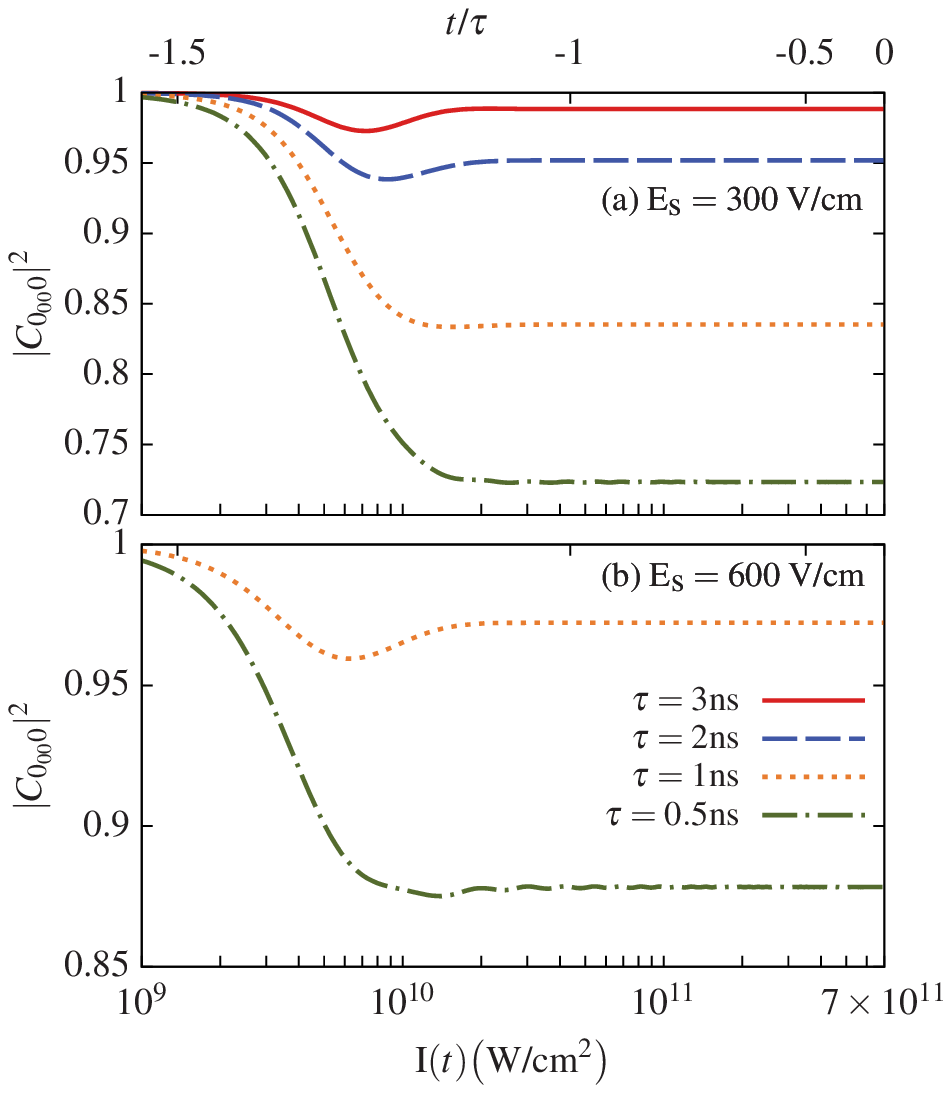}
  \caption{(Color online) For the state \pstate{0}{0}{0}{0}$_t$, 
we present the squares of the projection of the time-dependent
wave function onto the adiabatic ground state \ppstate{0}{0}{0}{0}
versus  the laser intensity $\textup{I}(t)$ 
for dc field strengths (a) $\Estatabs=\SI{300}{\fieldstrength}$ and (b) $\Estatabs= \SI{600}{\fieldstrength}$.
The Gaussian pulse has $\Ialign=\SI{7e11}{\intensity}$ and 
 $\tau=3$~ns (solid red), $\tau=2$~ns (blue long dashed), $\tau=1$~ns (orange dotted) 
and $\tau=0.5$~ns (green dashed-dotted). }
  \label{fig:function_fwhm_population_00_00_00_00}
\end{figure}

In \autoref{fig:function_fwhm_00_00_00_00}(a) and 
\autoref{fig:function_fwhm_00_00_00_00}(b)
we present 
the orientation $\left\langle\cos\theta\right\rangle$  of the ground
state \pstate{0}{0}{0}{0}$_t$ as a function of the laser intensity $\textup{I}(t)$
for  $\Estatabs=\SI{300}{\fieldstrength}$ and $\Estatabs=\SI{600}{\fieldstrength}$, respectively. 
The Gaussian pulses have $\Ialign=\SI{7e11}{\intensity}$  and several FWHM.
For comparison, the orientation of the adiabatic state \ppstate{0}{0}{0}{0} is also presented. 
For all these field-configurations, the orientation shows a
qualitatively similar behavior: 
 $\left\langle\cos\theta\right\rangle$ monotonically increases
as $\textup{I}(t)$ is increased, once the pendular limit is reached, 
the slope of  $\left\langle\cos\theta\right\rangle$ versus $\textup{I}(t)$ is reduced, and 
$\left\langle\cos\theta\right\rangle$ increases smoothly with a  value smaller than the 
adiabatic limit. This reduction on the final orientation is  
due to the non-adiabatic effects.
Analogously to the ground state of a linear molecule in combined electric and non-resonant laser 
fields, the loss of adiabaticity in the field-dressed dynamics of  \pstate{0}{0}{0}{0}$_t$ 
is due to the formation of the quasidegenerate pendular doublets~\cite{omiste:pra2012}. 
Thus, increasing the FWHM of the pulse will increase the orientation~\cite{nielsen:prl2012,omiste:pra2012,Sugawara2008,Muramatsu2009}. 
In \autoref{fig:detail_doublet_00_00_00_00}, we present how the pendular doublet between the
adiabatic states
  \ppstate{0}{0}{0}{0} and   \ppstate{1}{0}{1}{0} is formed.
To illustrate this loss of adiabaticity of  \pstate{0}{0}{0}{0}$_t$ we show the population 
of the adiabatic ground state  \ppstate{0}{0}{0}{0},   $|C_{0_{0,0}0}(t)|^2$,  in
\autoref{fig:function_fwhm_population_00_00_00_00}. Note that for the ground state,
$|C_{0_{0,0}0}(t)|^2+|C_{1_{0,1}0}(t)|^2=1$.
For $\Estatabs=\SI{300}{\fieldstrength}$, the rotational dynamics is adiabatic if the pulse has 
$\tau\ge 4$~ns, whereas for smaller values of $\tau$ the population transfer when the pendular doublet is formed could be very large. 
Using a $1$~ns pulse, the population of the adiabatic ground state is $|C_{0_{0,0}0}(t)|^2=0.835$ at
$t=0$, and due to the contribution of the anti-oriented state  \ppstate{1}{0}{1}{0}, 
 the orientation at $t=0$ of \pstate{0}{0}{0}{0}$_0$ is reduced to $\costheta=0.661$.
 By increasing the dc field strength to $\Estatabs=\SI{600}{\fieldstrength}$, the energy splitting of
the pendular doublets is increased, and a Gaussian pulse of  
 $\tau=2$~ns already gives rise to an adiabatic dynamics for \pstate{0}{0}{0}{0}$_t$. 
For a short pulse of $500$~ps, the ground state is still strongly oriented with 
$\costheta=0.747$ and the contribution of the adiabatic ground state at $t=0$
is   $|C_{0_{0,0}0}(0)|^2=0.878$. 
For both dc fields and $500$~ps, $\costheta$ shows an oscillatory behavior as $I(t)$ is increased,
which is due to the coupling between the two adiabatic states involved in the dynamics, \ie,
the mixing term     ${}_\text{p}\melement{0_{00}0}{\cos\theta}{1_{01}0}_\text{p}$. 

The lowest-lying  state in the irreducible representations with $M=1$,
\pstate{1}{0}{1}{1}$_t$, is relatively well separated of neighboring 
levels with the same symmetry. 
Thus, its field-dressed dynamics shows analogous features as 
those discussed above for  the \pstate{0}{0}{0}{0}$_t$, and the formation of the pendular 
pair is the only  effect provoking the loss of adiabaticity. 
Indeed, for $\Estatabs=\SI{300}{\fieldstrength}$, a
Gaussian pulse with   $\Ialign=\SI{7e11}{\intensity}$ and 
$\tau=5$~ns  gives rise to an adiabatic dynamics for this state.

\subsection{Dynamics of the state  \pstate{3}{0}{3}{3}$_t$ }
\begin{figure}[htb]
  \centering
  \includegraphics[width=.48\textwidth]{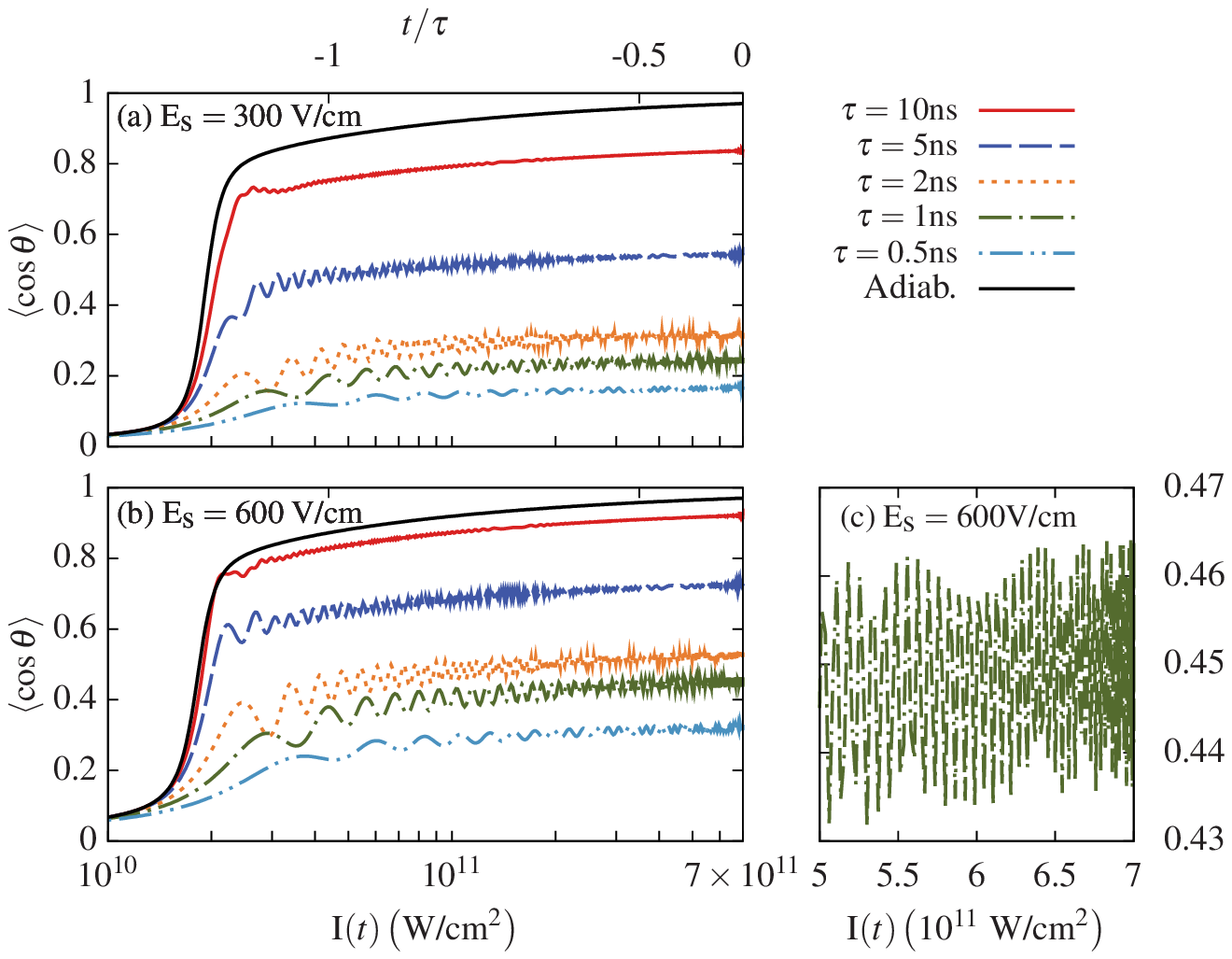}
  \caption{(Color online) For the state \pstate{3}{0}{3}{3}$_t$,
we present the orientation cosine $\left\langle\cos\theta\right\rangle$ as a function of $\textup{I}(t)$ for 
(a) $\Estatabs=\SI{300}{\fieldstrength}$ and (b) $\SI{600}{\fieldstrength}$.
A detail of the oscillations of  $\left\langle\cos\theta\right\rangle$ for
  $\SI{600}{\fieldstrength}$
is shown in panel (c).
The peak intensity of the pulses is $\Ialign=\SI{7e11}{\intensity}$, and the FWHM are
$\tau=10$~ns (red solid line), $\tau=5$~ns (dark blue long dashed line), $\tau=2$~ns (orange dotted line), $\tau=1$~ns (dark olive
 green dot-dashed line), $\tau=0.5$~ns (light blue double-dot-dashed  line). 
The adiabatic results are also shown (black solid line).}
  \label{fig:function_fwhm_03_00_03_03}
\end{figure}
For higher values of $M$, 
the avoided crossings leave their fingerprints in 
the field-dressed dynamics of  the corresponding ground state. 
As an example, we show in \autoref{fig:function_fwhm_03_00_03_03} 
the orientation  cosine of  \pstate{3}{0}{3}{3}$_t$ for 
$\Estatabs=\SI{300}{\fieldstrength}$ and $\Estatabs=\SI{600}{\fieldstrength}$,
the adiabatic value of $\costheta$ is also included. 
Comparing these results to those  of the absolute ground state  \pstate{0}{0}{0}{0}$_t$
in \autoref{fig:function_fwhm_00_00_00_00}, 
 two main differences are encountered.
First,  $\costheta$ initially increases as  $\textup{I}(t)$ is increased, and 
once the pendular doublet is formed $\costheta$ oscillates around a mean value. 
Second, even the $10$~ns Gaussian pulses do not  to ensure an adiabatic dynamics
for both dc field strengths. For instance, at $t=0$ 
we obtain  $\costheta=0.921$ for $\Estatabs=\SI{600}{\fieldstrength}$ and $\tau=10$~ns,
 whereas the adiabatic value is $\costheta=0.970$. 
As the FWHM is increased, the amplitude of the oscillations 
of $\costheta$ is reduced.

\begin{figure}[t]
  \centering 
 \includegraphics[width=.48\textwidth]{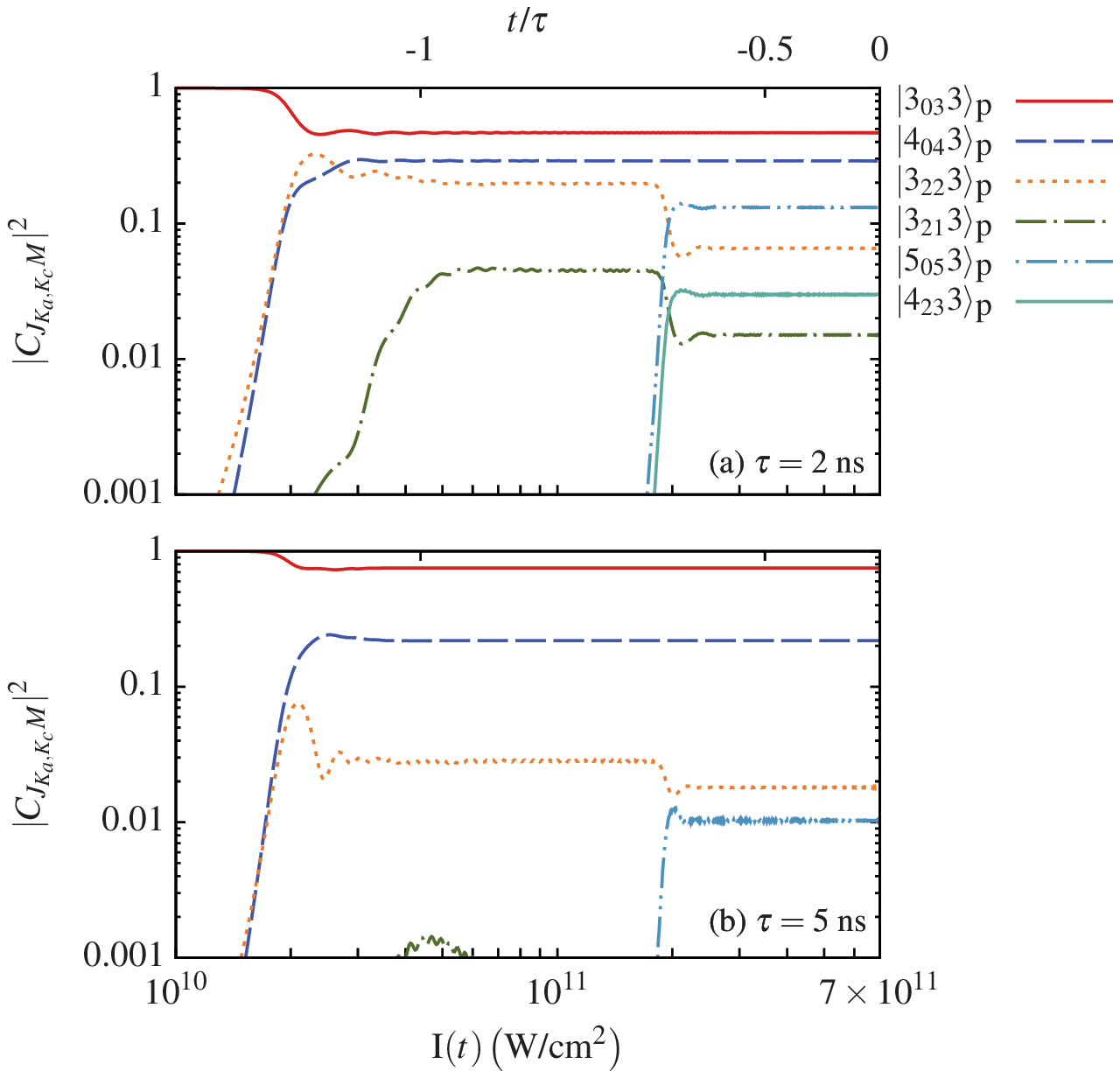}
    \caption{(Color online) 
   For  the state \pstate{3}{0}{3}{3}$_t$, 
we present the squares of the projections of the time dependent
wave function onto the adiabatic states 
versus  the laser intensity $\textup{I}(t)$. 
The Gaussian pulse has $\Ialign=\SI{7e11}{\intensity}$ and (a) $\tau=5$~ns and (b) $\tau=2$~ns,
and the dc field strength is $\Estatabs=\SI{300}{\fieldstrength}$.}
  \label{fig:population_2_5_fwhm_03_00_03_03}
\end{figure}

\begin{figure}[t]
  \centering 
\includegraphics[width=.48\textwidth,angle=0]{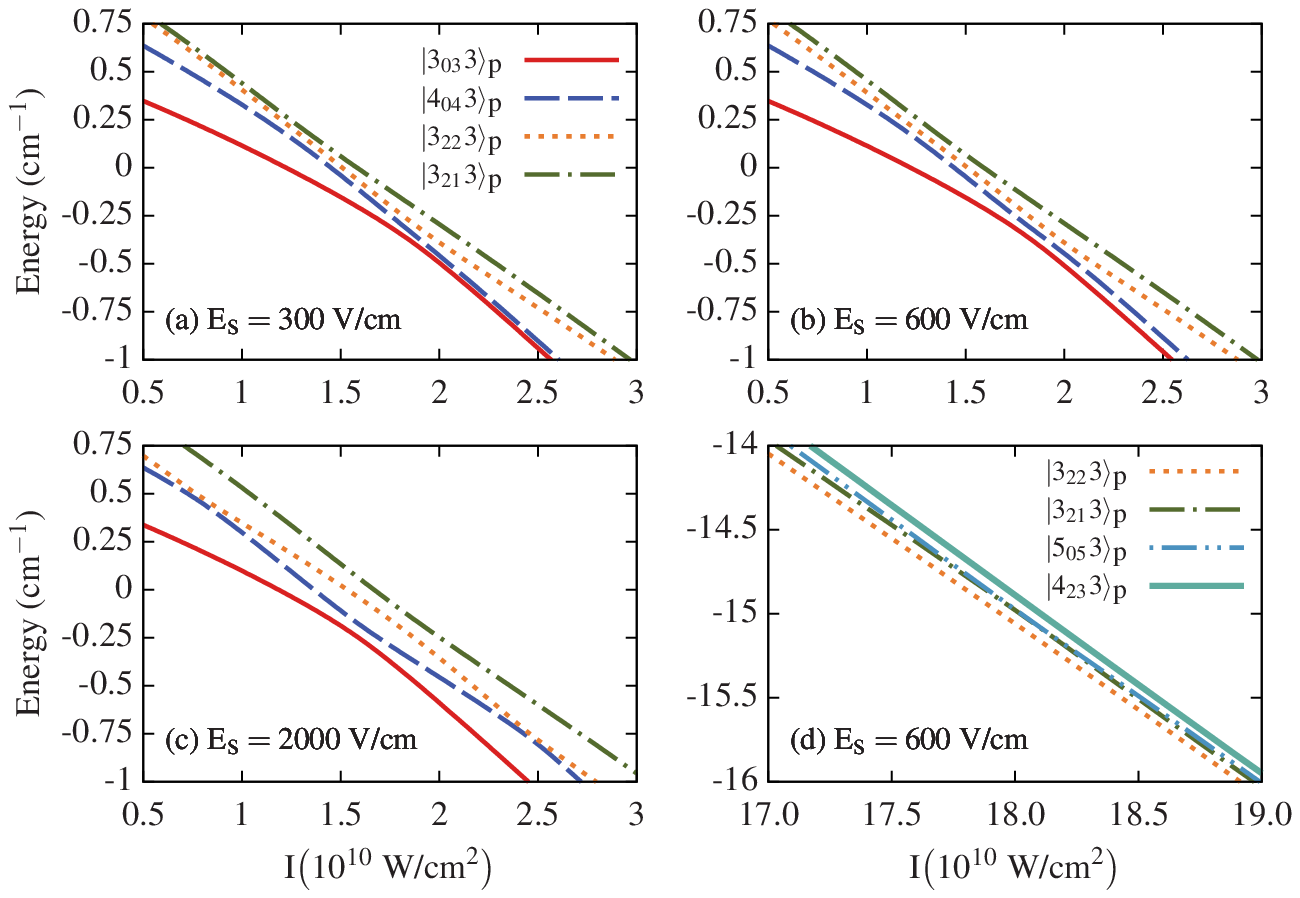}
    \caption{(Color online) 
(a), (b) and  (c)  Adiabatic energy structure when 
the pendular doublets between the states \ppstate{3}{0}{3}{3} and \ppstate{4}{0}{4}{3} is formed 
for  $\Estatabs=\SI{300}{\fieldstrength}$,  $\Estatabs=\SI{600}{\fieldstrength}$ and
 $\Estatabs=\SI{2}{\kfieldstrength}$, respectively. (d) Avoided crossing between the adiabatic states
\ppstate{3}{2}{2}{3}, \ppstate{3}{2}{1}{3},  \ppstate{5}{0}{5}{3} and  \ppstate{4}{2}{3}{3}.
  \label{fig:avoided_crossing_detail_03_00_03_03}}
\end{figure}

The oscillations in the evolution of  $\costheta$ can be 
explained due to the coupling between the adiabatic states contributing to the dynamics.
The presence of  avoided crossings in the spectrum provokes
that  adiabatic states from different pendular doublets are populated during the
rotational dynamics of \pstate{3}{0}{3}{3}$_t$.  
In~\autoref{fig:population_2_5_fwhm_03_00_03_03}(a)
and ~\autoref{fig:population_2_5_fwhm_03_00_03_03}(b), we plot 
the population  of the pendular adiabatic states 
for  $\Estatabs=\SI{300}{\fieldstrength}$, $\Ialign=\SI{7e11}{\intensity}$ and 
$\tau=2$~ns and $5$~ns, respectively. 
 We start analyzing in detail the results for $\tau=2$~ns. 
To rationalize the population redistribution taking place around  $\textup{I}(t)\approx \SI{2e10}{\intensity}$,
see \autoref{fig:population_2_5_fwhm_03_00_03_03}(a), 
we present a detail of the adiabatic level structure 
in \autoref{fig:avoided_crossing_detail_03_00_03_03}(a).
As the pendular doublet between the adiabatic levels  \ppstate{3}{0}{3}{3} and \ppstate{4}{0}{4}{3}
is formed,
\ppstate{4}{0}{4}{3} suffers an avoided crossing with \ppstate{3}{2}{2}{3}.
For the states \ppstate{3}{0}{3}{3} and \ppstate{4}{0}{4}{3} in the pendular doublet, the maximum value of 
adiabatic parameter is  $\eta\approx 1.3$.
The maximum of $\eta$ for the state \ppstate{3}{2}{2}{3} with 
\ppstate{3}{0}{3}{3} and \ppstate{4}{0}{4}{3} are $\eta\approx0.44$ and $0.46$ respectively.
Thus, the two oriented states  \ppstate{3}{0}{3}{3} and \ppstate{3}{2}{2}{3} are also coupled, 
even if their energy separation is larger.
These values of $\eta$ indicate that the dynamics in this region is not adiabatic.
Indeed,  the population of 
these two states  \ppstate{4}{0}{4}{3}  and  \ppstate{3}{2}{2}{3}  simultaneously increases
as  the one of  \ppstate{3}{0}{3}{3} decreases. 
Once the pendular pair between  \ppstate{3}{0}{3}{3} and \ppstate{4}{0}{4}{3} is formed their 
populations keep a constant behavior as $\textup{I}(t)$ is increased.
We observe the formation of the second pendular doublet in this irreducible 
representation for $\textup{I}(t)\approx\SI{3e10}{\intensity}$: 
the population of the  adiabatic states  \ppstate{3}{2}{2}{3} and  \ppstate{3}{2}{1}{3}
decreases and increases, respectively. 
By further increasing $\textup{I}(t)$ the states \ppstate{3}{2}{2}{3} and  \ppstate{3}{2}{1}{3}
suffer an avoided crossing with those from the third pendular doublet 
 \ppstate{5}{0}{5}{3} and  \ppstate{4}{2}{3}{3},
 see \autoref{fig:avoided_crossing_detail_03_00_03_03}(d).  
 Through this avoided crossing, there is a
 strong coupling between the  oriented states 
 \ppstate{3}{2}{2}{3} and    \ppstate{5}{0}{5}{3}
with the adiabatic parameter reaching the maximum value $\eta\approx 1.93$,  and between the antioriented  ones
 \ppstate{3}{2}{1}{3}  and  \ppstate{4}{2}{3}{3} with $\eta\approx 1.92$.
In both cases the dynamics is not adiabatic, and 
we observe in 
 \autoref{fig:population_2_5_fwhm_03_00_03_03}(a) how their populations are interchanged
 around $\textup{I}(t)\approx \SI{2e11}{\intensity}$.
The oscillations  in $\costheta$ are due to the coupling between all these adiabatic states that are
populated. 
As the laser intensity is increased, these levels achieve the  pendular regime and these crossed 
matrix elements between states in the same pendular pair, \ie,
${}_\text{p}\melement{3_{03}3}{\cos\theta}{4_{04}3}_\text{p}$,
${}_\text{p}\melement{3_{22}3}{\cos\theta}{3_{21}3}_\text{p}$ and
${}_\text{p}\melement{5_{05}3}{\cos\theta}{4_{23}3}_\text{p}$, 
approach zero. 
As $t$ increases,  the frequency of the oscillation varies because different pendular
adiabatic states dominate the field-dressed dynamics,
see  \autoref{fig:function_fwhm_03_00_03_03}(c). 
At  $t=0$, the field-dressed wave function of the state \pstate{3}{0}{3}{3}$_0$ has significant 
contributions from $6$ different adiabatic states, which gives rise to a weak orientation
 $\costheta=0.327$. 
For $\tau=5$~ns, the dynamics is more adiabatic. Thus, when the pendular doublets are formed 
the interchange of  population is smaller than for a $\tau=2$~ns 
pulse, see  \autoref{fig:population_2_5_fwhm_03_00_03_03}(b). 
The avoided crossing are not crossed adiabatically. 
Indeed, the population of the states   \ppstate{3}{2}{1}{3} and  \ppstate{4}{2}{3}{3} is smaller than
$0.001$, and the field-dressed dynamics of  \pstate{3}{0}{3}{3}$_t$ is dominated by
the adiabatic states \ppstate{3}{0}{3}{3}, \ppstate{4}{0}{4}{3}, \ppstate{3}{2}{2}{3}
and  \ppstate{5}{0}{5}{3}. This explain that the oscillations of $\costheta$ are reduced, and that
at $t=0$ this state shows a significant orientation with $\costheta=0.547$.

For other ground states, such as  \pstate{2}{0}{2}{2}$_t$ and \pstate{4}{0}{4}{4}$_t$, 
we have encountered similar phenomena, and their rotational dynamics is strongly dominated by
avoided crossings.

\subsection{Influence of FWHM of the Gaussian pulse}
\begin{figure}[t]
  \centering
  \includegraphics[width=.32\textwidth]{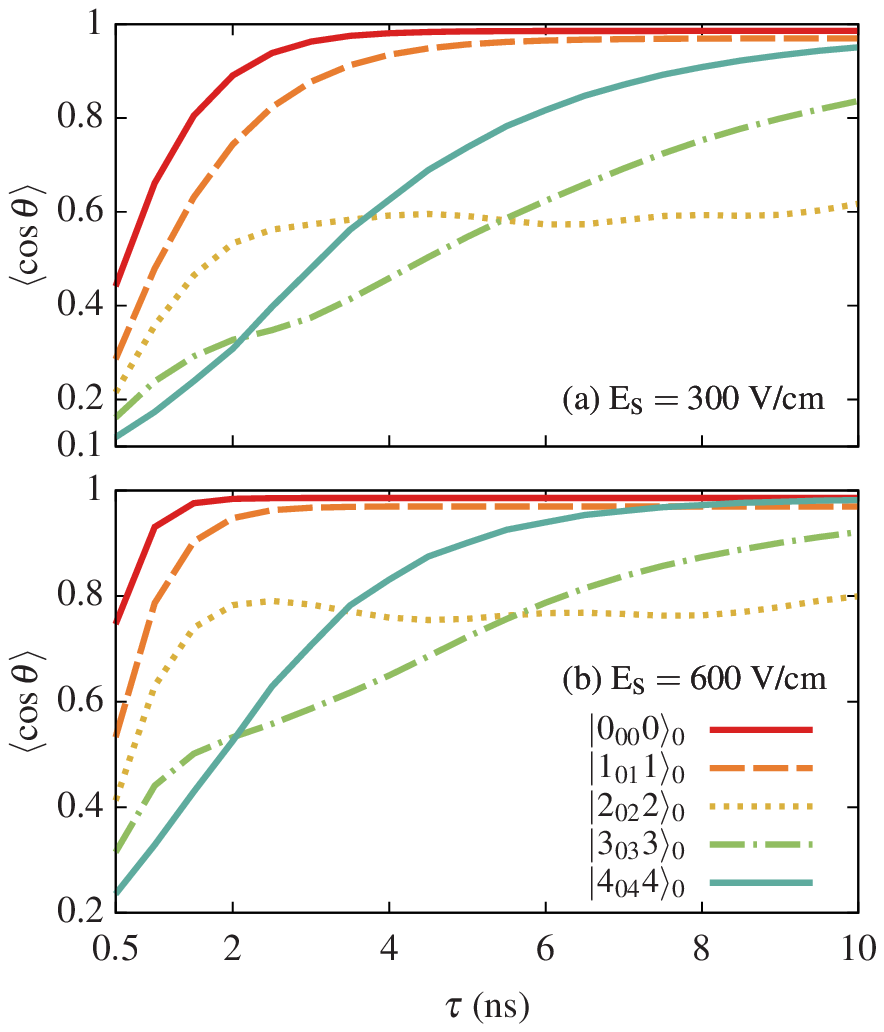}
  \caption{(Color online) 
For the ground states of several irreducible representations, we plot the expectation value 
$\left\langle\cos\theta\right\rangle$ at $t=0$ as a function of the FWHM of the Gaussian pulse
for  (a) $\Estatabs=\SI{300}{\fieldstrength}$ and (b) $\Estatabs=\SI{600}{\fieldstrength}$.
The peak intensity is fixed to $\Ialign=\SI{7e11}{\intensity}$.}
  \label{fig:function_fwhm_final}
\end{figure}

We consider now the ground states of the irreducible representations with $M\le 4$, that is the
states \pstate{0}{0}{0}{0}$_t$, \pstate{1}{0}{1}{1}$_t$, \pstate{2}{0}{2}{2}$_t$, \pstate{3}{0}{3}{3}$_t$   
and \pstate{4}{0}{4}{4}$_t$.
In this section we investigate the impact of the temporal width of the Gaussian pulse 
on their rotational dynamics.
Their orientation at the peak intensity, \ie, at $t=0$, 
are plotted  versus $\tau$  
in \autoref{fig:function_fwhm_final}(a) and \autoref{fig:function_fwhm_final}(b)
for $\Estatabs=\SI{300}{\fieldstrength}$ and $\Estatabs=\SI{600}{\fieldstrength}$, respectively.
The peak intensity is fixed to $\Ialign=\SI{7e11}{\intensity}$. 

In contrast to the mixed-field orientation of a linear molecules~\cite{omiste:pra2012}, 
a smaller field-free rotational energy does not imply a larger orientation. 
For instance, using $\Estatabs=\SI{300}{\fieldstrength}$, 
the state \pstate{4}{0}{4}{4}$_0$ shows a larger orientation than  
\pstate{2}{0}{2}{2}$_0$ and \pstate{3}{0}{3}{3}$_0$ for $\tau\gtrsim 2$~ns  and $3.5$~ns,
respectively.
For $\tau\gtrsim 6$~ns, the state  \pstate{2}{0}{2}{2}$_0$ is the least oriented.  
As indicated above, the non-adiabatic features of the field-dressed dynamics are due to the
formation of the pendular doublets and to the avoided crossings. 
By increasing $\tau$, we can ensure that less population is transferred from the oriented state to
the antioriented one or vice-versa as the pendular doublet is formed.
However, the characteristic time scale of the avoided crossings is different, and
significantly larger FWHM are needed to pass them adiabatically. 
Each irreducible representation is characterized by a certain field-dressed level structure, and,
therefore, by an amount of avoided crossings which contribute to the complexity of the 
rotational dynamics. 
The absence of avoided crossings close to the ground states gives rise
to a monotonic increase of $\costheta$ approaching the adiabatic limit as $\tau$ is increased,
this behavior is observed in the levels \pstate{0}{0}{0}{0}$_t$ and \pstate{1}{0}{1}{1}$_t$.
In contrast, the influence of an avoided crossing in the dynamics implies that 
significantly longer pulses are needed to reach the adiabatic limit, \eg,
the levels  \pstate{2}{0}{2}{2}$_t$  and  \pstate{3}{0}{3}{3}$_t$.
In particular, for $\tau\gtrsim 2$~ns, 
the orientation  $\costheta$ of \pstate{2}{0}{2}{2}$_t$ shows a smooth oscillatory behavior as a function of $\tau$.

\subsection{Influence of electric-field strength}

\begin{figure}[t]
  \centering
  \includegraphics[width=.32\textwidth]{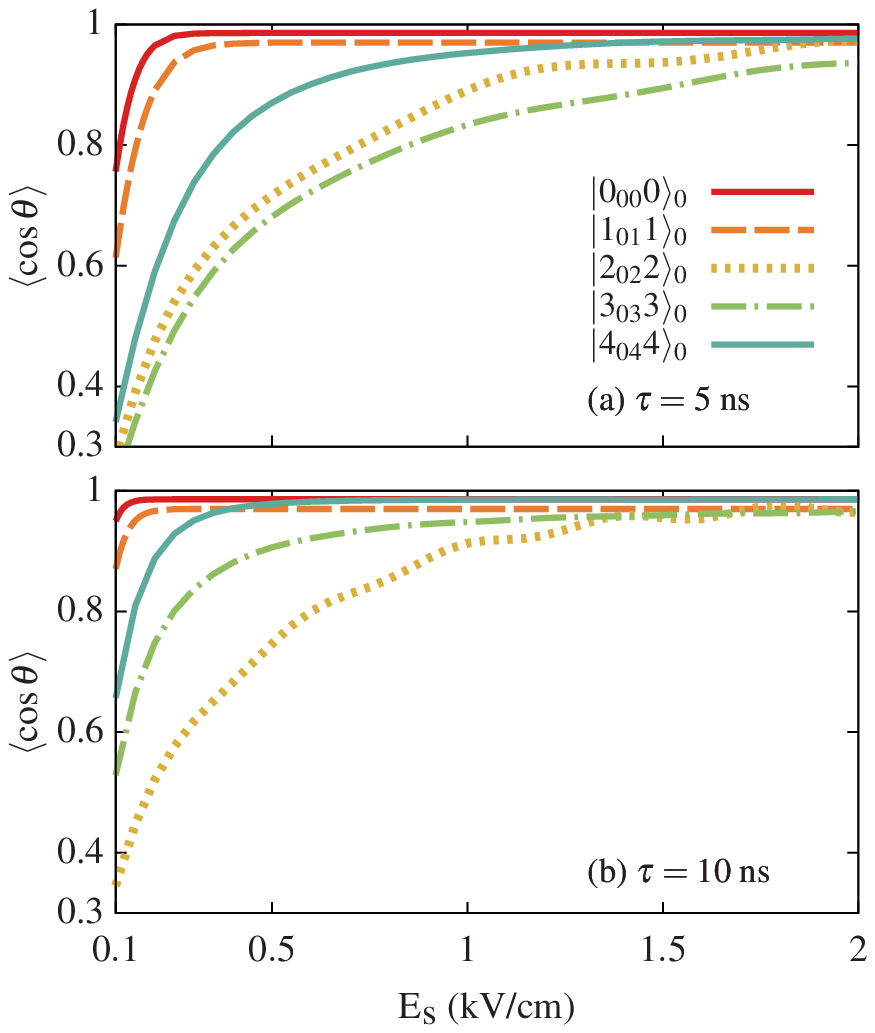}
  \caption{(Color online) For the ground states of several irreducible representations, 
we present their expectation value 
$\left\langle\cos\theta\right\rangle$ once the peak intensity is reached at 
$t=0$ as a function of the dc field strength $\Estatabs$
for  (a) $\tau=5$~ns and (b) $\tau=10$~ns.
The peak intensity is fixed to $\Ialign=\SI{7e11}{\intensity}$.}
  \label{fig:function_es}
\end{figure}
In this section, we consider the same set of ground states  and analyze their orientation
at $t=0$ as a function of the dc field strength, see \autoref{fig:function_es}. 
For a given laser pulse, the largest
is the energy gap between the two levels in a pendular pair,  the less efficient is the population 
transfer when the doublet is formed~\cite{omiste:pra2012}. This statement also holds for 
asymmetric top molecules. However, the impact of the electric field on the avoided crossings will
influence the rotational dynamics.
For both pulses, we encounter that the states  
\pstate{2}{0}{2}{2}$_0$  and \pstate{3}{0}{3}{3}$_0$ 
present a smaller orientation than
\pstate{4}{0}{4}{4}$_0$. 
The level \pstate{3}{0}{3}{3}$_0$  is more (less) oriented than
\pstate{2}{0}{2}{2}$_0$ for the $\tau=10$~ns ($\tau=5$~ns) pulses. 
Our calculations show that for these ground states, the adiabatic pendular limit could be
reached using a strong electric field and a $10$~ns pulse, see \autoref{fig:function_es}~(b). 
 
 \section{Field-dressed dynamics of  excited states}
\label{se:excited_states}

In this section we analyze the mixed-field dynamics of
two excited rotational states from different irreducible representations. 
The field-dressed spectrum is characterized by a high  density of adiabatic states 
and a large number of avoided crossings between neighboring levels. As a consequence,
the rotational dynamics is more complex, and it is harder to achieve the diabatic limit.

\subsection{Dynamics of the state  \pstate{4}{0}{4}{3}$_t$ }
\begin{figure}[t]
  \centering
  \includegraphics[width=.48\textwidth]{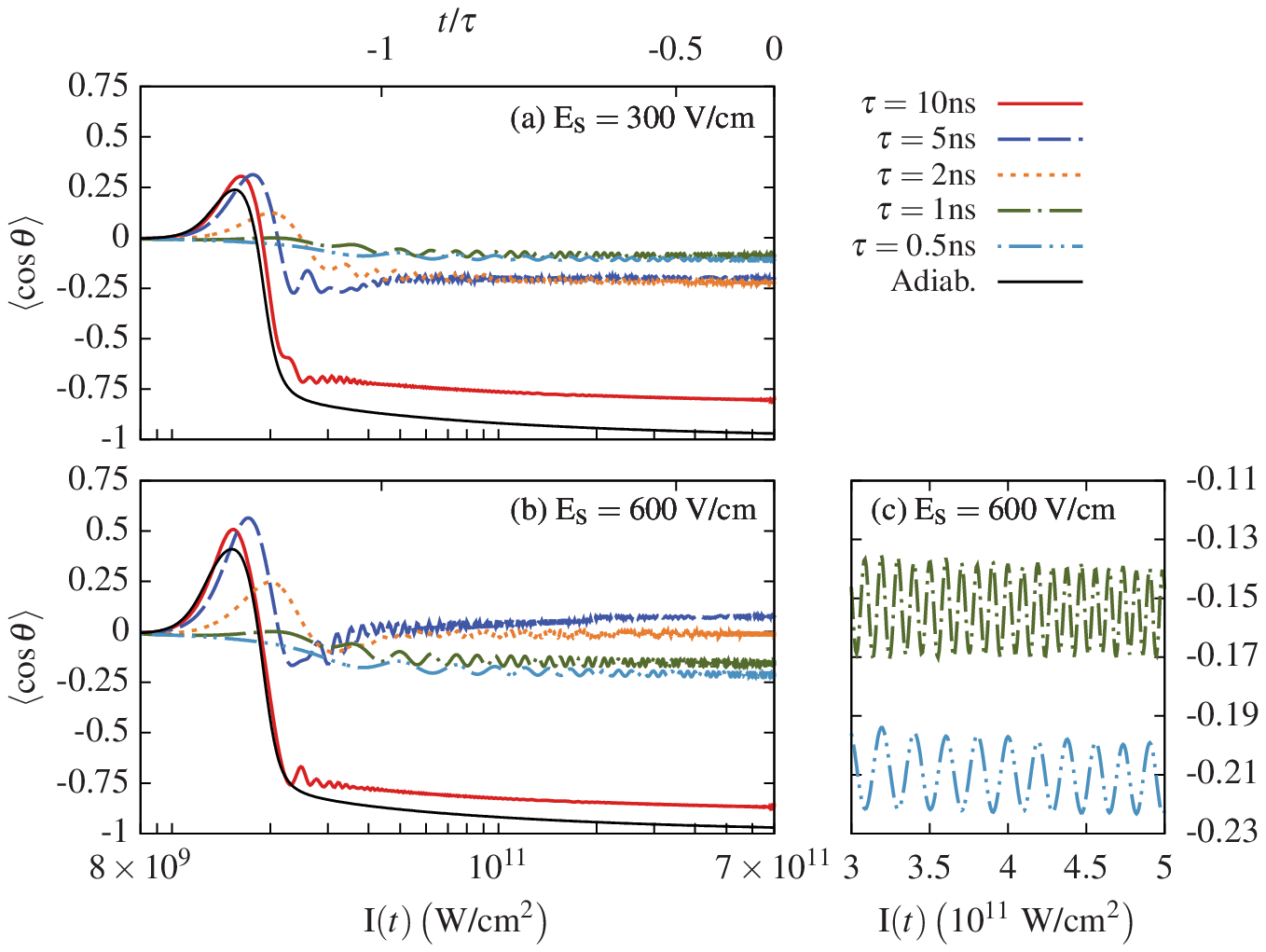}
  \caption{(Color online) For the state \pstate{4}{0}{4}{3}$_t$,
we present  $\left\langle\cos\theta\right\rangle$ versus $\textup{I}(t)$ for 
(a)         $\Estatabs=\SI{300}{\fieldstrength}$ and (b) $\SI{600}{\fieldstrength}$.
A detail of the oscillations of  $\left\langle\cos\theta\right\rangle$ for
 $\SI{600}{\fieldstrength}$ is shown in panel (c).
The peak intensity of the pulses is $\Ialign=\SI{7e11}{\intensity}$, and the FWHM are
$\tau=10$~ns (red solid line), $5$~ns (dark blue long dashed line), $2$~ns (orange dotted line),
  $1$~ns (dark olive  green dot-dashed line), $0.5$~ns (light blue double-dot-dashed line). The
  adiabatic results is also  plotted (black solid line).}
  \label{fig:function_fwhm_04_00_04_03}
\end{figure}
\begin{figure}[t]
  \centering 
\includegraphics[width=.48\textwidth]{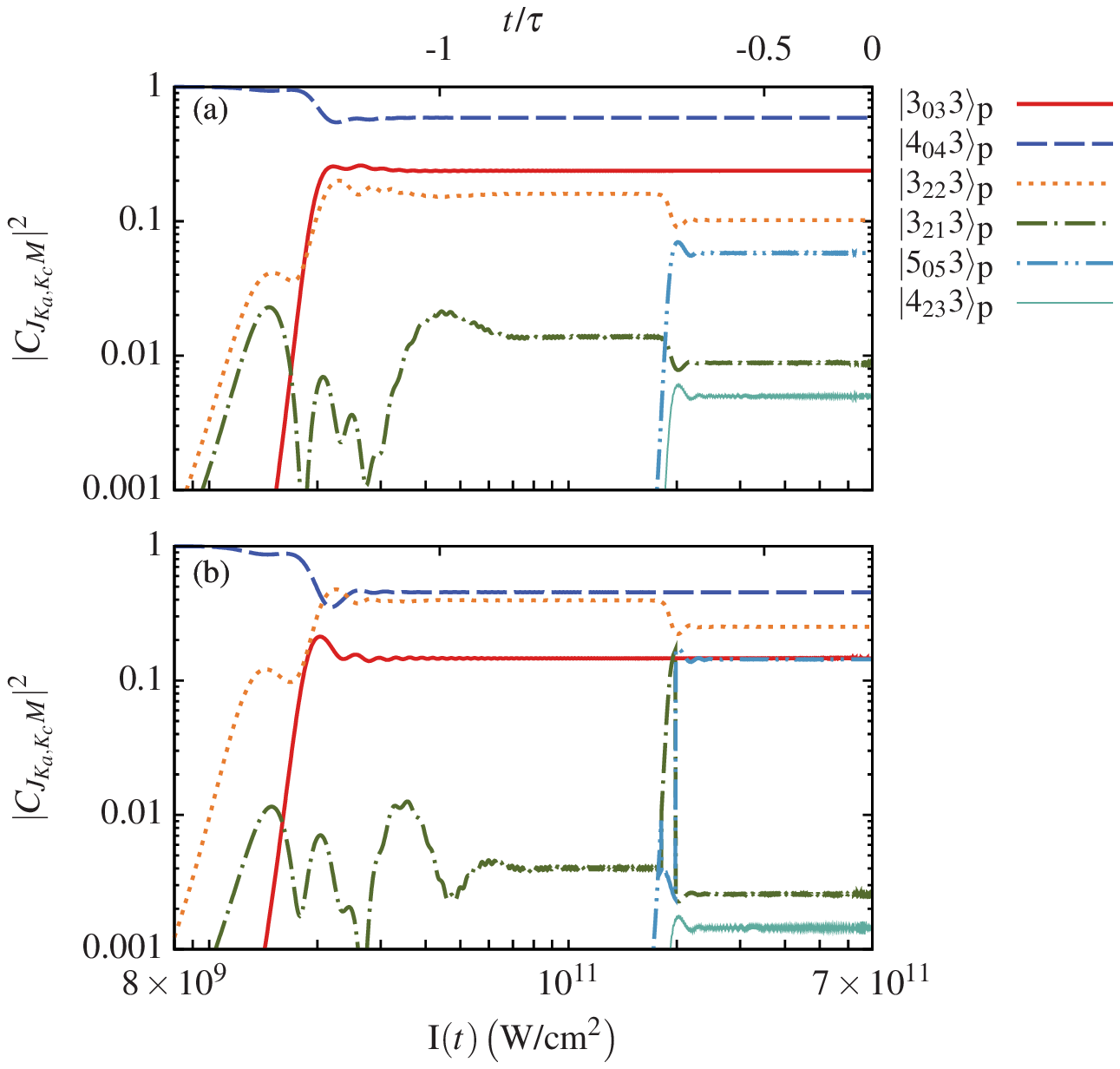}
    \caption{(Color online) 
   For  the state \pstate{4}{0}{4}{3}$_t$, 
we present the squares of the projections of the time dependent
wave function onto several adiabatic states 
versus  the laser intensity $\textup{I}(t)$ for
 dc field strengths (a) $\Estatabs=\SI{300}{\fieldstrength}$ and 
 (b) $\Estatabs=\SI{600}{\fieldstrength}$.
The Gaussian pulse has $\Ialign=\SI{7e11}{\intensity}$ and $\tau=5$~ns.
The notation and label of the adiabatic states is the same as in \autoref{fig:population_2_5_fwhm_03_00_03_03}.}
  \label{fig:population_5_fwhm_04_0_4_03}
\end{figure}

As a first example, we investigate the dynamics of the rotational
state \pstate{4}{0}{4}{3}$_t$ which forms the  pendular pair with \pstate{3}{0}{3}{3}$_t$. 
In \autoref{fig:function_fwhm_04_00_04_03}(a) and \autoref{fig:function_fwhm_04_00_04_03}(b) the
orientation of \pstate{4}{0}{4}{3}$_t$  is plotted versus $\textup{I}(t)$ for  
(a) $\Estatabs=\SI{300}{\fieldstrength}$ and (b) $\Estatabs=\SI{600}{\fieldstrength}$, respectively.
In contrast to the case of a linear molecule in parallel dc and ac fields, this state does not
show the same orientation but in opposite direction as its partner in the pendular doublet the
level \pstate{3}{0}{3}{3}$_t$, except if the dynamics is adiabatic or very close to it. 
This can be explained in terms of the avoided crossings,
which affect in different ways the rotational dynamics of \pstate{3}{0}{3}{3}$_t$ and 
\pstate{4}{0}{4}{4}$_t$.  
We present in 
\autoref{fig:population_5_fwhm_04_0_4_03}(a)
and
\autoref{fig:population_5_fwhm_04_0_4_03}(b)
the contributions of adiabatic states to the rotational dynamics of \pstate{4}{0}{4}{3}$_t$    
for a $5$~ns pulse with
$\Estatabs=\SI{300}{\fieldstrength}$ and $\SI{600}{\fieldstrength}$, respectively.
For both dc field strengths, we observe that the population redistribution starts first to the adiabatic
states
\ppstate{3}{2}{2}{3} and \ppstate{3}{2}{1}{3}.
Due to the rotational constants of BN, the states  \ppstate{3}{2}{2}{3} and  \ppstate{3}{2}{1}{3}
are quasi-degenerate in energy in the absence of the fields, and as $\textup{I}(t)$ varies they posses a 
very close energy, see ~\autoref{fig:avoided_crossing_detail_03_00_03_03}(a) and
\autoref{fig:avoided_crossing_detail_03_00_03_03}(b). 
As a consequence, population is initially transferred to both levels, but more to
\ppstate{3}{2}{2}{3} which lies  closer to \ppstate{4}{0}{4}{3}. 
For a stronger laser intensity, the pendular pair \ppstate{3}{0}{3}{3} and \ppstate{4}{0}{4}{3} starts to form
and $|C_{3_{03}3}(t)|^2$ increases.
The main difference between both dc field strengths is that
for $\Estatabs=\SI{300}{\fieldstrength}$, \ppstate{3}{0}{3}{3} acquires the largest population;
whereas for  $\Estatabs=\SI{600}{\fieldstrength}$ is \ppstate{3}{2}{2}{3}. 
Indeed, at the
largest dc field the avoided crossing is passed less adiabatically, and more population is
transferred to the  \ppstate{3}{2}{2}{3}, because the coupling between the states is larger.
Whereas for  $\Estatabs=\SI{600}{\fieldstrength}$, the pendular
pair among \ppstate{3}{0}{3}{3} and \ppstate{4}{0}{4}{3} 
is formed more adiabatically because the energy splitting in the doublets is larger, 
and  \ppstate{3}{0}{3}{3} is less populated than for  $\Estatabs=\SI{300}{\fieldstrength}$.
In these plots, we also observe how the second pendular doublet between \ppstate{3}{2}{2}{3}
and \ppstate{3}{2}{1}{3} is formed around $\textup{I}(t)\approx\SI{2.3e10}{\intensity}$. 
 For stronger laser intensities, the next pendular pair 
 \ppstate{5}{0}{5}{3} and  \ppstate{4}{2}{3}{3} is also populated due to the avoided crossing
that these levels suffer with those forming the second doublet, see \autoref{fig:avoided_crossing_detail_03_00_03_03}(d).
The couplings between these six pendular states provoke the oscillatory behavior of $\costheta$. 
Let us mention that the first avoided crossing is not crossed adiabatically using a $10$~ns
pulse, but the population of the state \ppstate{3}{2}{2}{3} is smaller than $0.03$ at $t=0$ for both field strengths. 
Only these $10$~ns pulses give rise to a significant antiorientation with values close to
the adiabatic predictions.

\begin{figure}[t]
  \centering 
\includegraphics[width=.4\textwidth,angle=0]{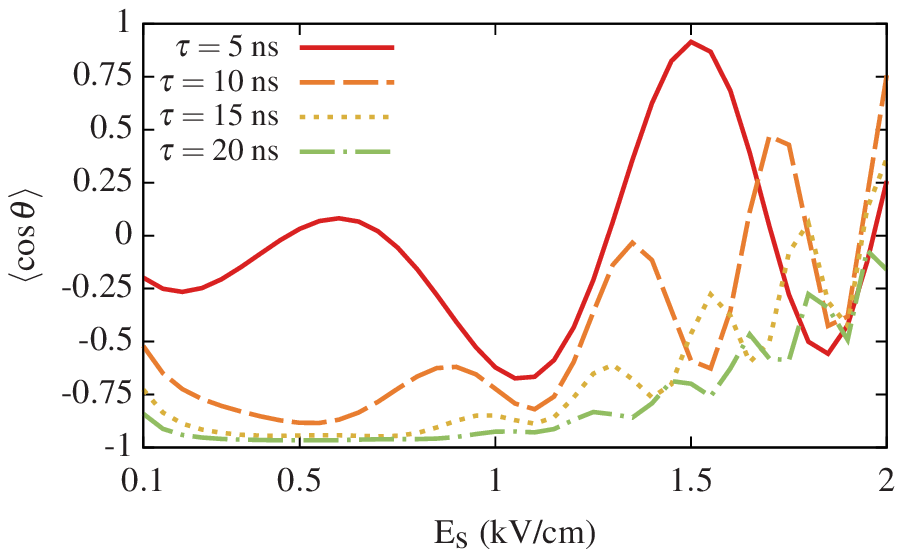}
    \caption{(Color online) 
   For  the state \pstate{4}{0}{4}{3}$_0$, 
we show the orientation at $t=0$ versus the electric field strength $\Estatabs$, for
a laser pulse with $\Ialign=\SI{7e11}{\intensity}$ and several FWHM.}
  \label{fig:state_4043_versus_es}
\end{figure}

We investigate now the rotational dynamics of this state  as the dc 
field is increased.
For four 
Gaussian pulses with $\Ialign=\SI{7e11}{\intensity}$, the orientation of 
\pstate{4}{0}{4}{3}$_0$ at $t=0$ is plotted versus $\Estatabs$ 
in \autoref{fig:state_4043_versus_es}. 
For this state, the adiabatic prediction is $\costheta=-0.970$, which is independent of $\Estatabs$.
Our time-dependent calculations show that the orientation oscillates as $\Estatabs$ in increased.
This behavior can be explained in terms of the avoided crossings, and their evolution as 
$\Estatabs$ varies. 
At $\Estatabs=\SI{300}{\fieldstrength}$, the pendular states
\ppstate{4}{0}{4}{3} and \ppstate{3}{2}{2}{3} suffer an avoided crossing for
$\textup{I}(t)\approx\SI{1.53e10}{\intensity}$, before the pendular
doublet  \ppstate{3}{0}{3}{3}, \ppstate{4}{0}{4}{3} is formed, see \autoref{fig:avoided_crossing_detail_03_00_03_03}(a).
By increasing  $\Estatabs$, this avoided crossing  is split into two, \eg,
for  $\Estatabs=\SI{2}{\kfieldstrength}$, the first and second one appear at 
$\textup{I}(t)\approx\SI{7.8e9}{\intensity}$ and $\textup{I}(t)\approx\SI{2.1e10}{\intensity}$, respectively, 
and the minimal energy between the two states in the
pendular doublet is reached at  $\textup{I}(t)\approx\SI{2.47e10}{\intensity}$, see 
\autoref{fig:avoided_crossing_detail_03_00_03_03}(c). 
The rotational dynamics through these avoided crossings as $\textup{I}(t)$ is varied strongly depends on the
FWHM of the laser pulse. 
For instance, using a $10$~ns pulse and $\Estatabs=\SI{2}{\kfieldstrength}$,
these two avoided crossings are passed non-adiabatically and the  pendular level  \ppstate{3}{2}{2}{3} is 
populated, and the formation of the two pendular doublets 
\ppstate{3}{0}{3}{3},\ppstate{4}{0}{4}{3} 
and
\ppstate{3}{2}{2}{3},\ppstate{3}{2}{1}{3}
is also non-adiabatic. 
Furthermore, the next avoided crossing between the pendular states
\ppstate{3}{2}{1}{3} and \ppstate{5}{0}{5}{3},
see \autoref{fig:avoided_crossing_detail_03_00_03_03}(d), is also crossed non-adiabatically.  
Hence, the final orientation of \pstate{4}{0}{4}{3}$_0$
strongly depends on the rotational dynamics through these avoided crossings. 
Indeed,  as  $\Estatabs$ is increased the population redistribution through these
avoided crossings is increased; whereas less population is transferred when the pendular doublets
are formed as occurs in linear molecules. 
For this state, to reach an adiabatic dynamics through the avoided crossings, longer laser pulses are needed, 
but the dc field  should  be chosen properly.  
For instance, a $20$~ns pulse ensures an adiabatic dynamics of this state with
$\SI{300}{\fieldstrength}\lesssim\Estatabs\lesssim\SI{1}{\kfieldstrength}$,
whereas for stronger dc fields, it is still non-adiabatic. 

\subsection{Dynamics of the state  \pstate{3}{0}{3}{1}$_t$ }
\begin{figure}[t]
  \centering
 \includegraphics[width=.4\textwidth]{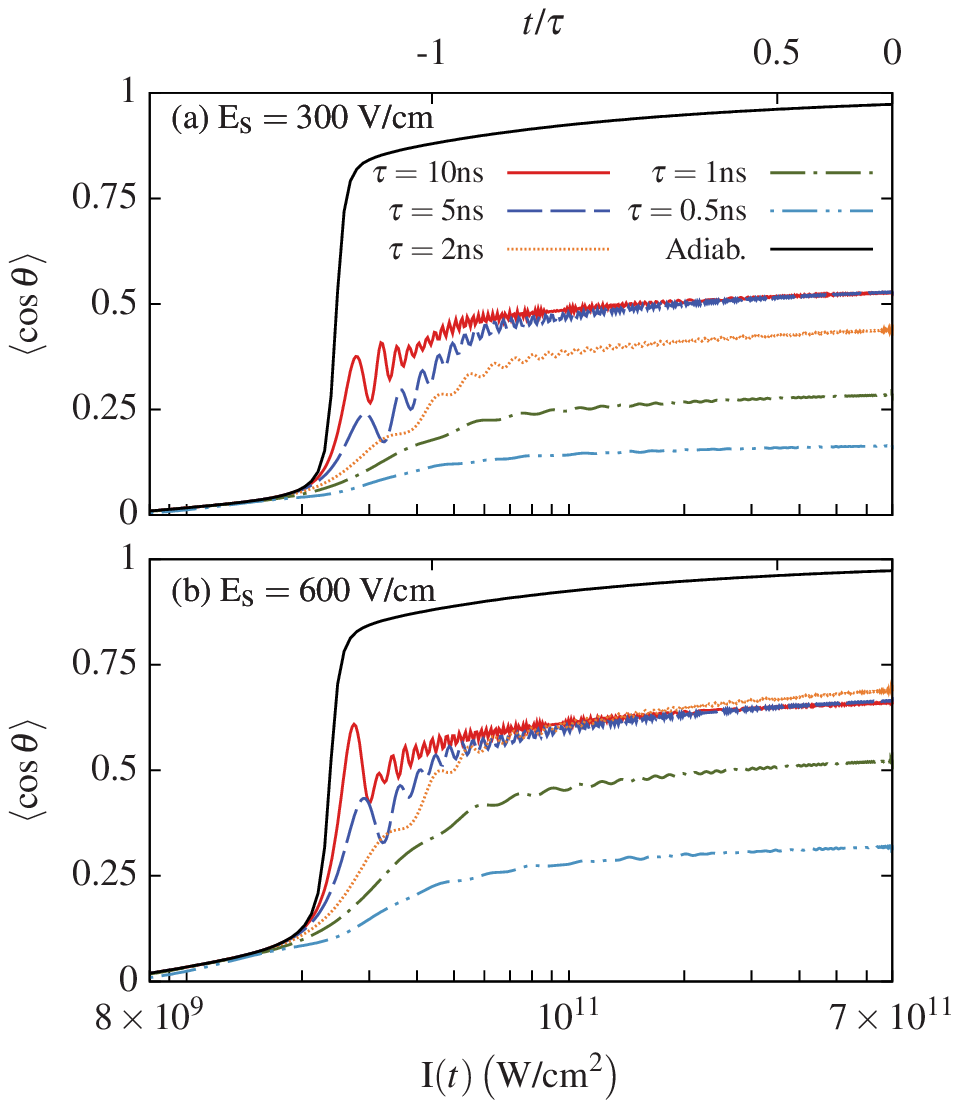}
  \caption{(Color online)
  For  the state \pstate{3}{0}{3}{1}$_t$, 
we plot $\left\langle\cos\theta\right\rangle$ versus $\textup{I}(t)$ for 
(a) $\Estatabs=\SI{300}{\fieldstrength}$ and (b) $\Estatabs=\SI{600}{\fieldstrength}$.
The peak intensity of the pulses is $\Ialign=\SI{7e11}{\intensity}$, and the FWHM are
$\tau=10$~ns (red solid line), $\tau=5$~ns (dark blue long dashed line), $\tau=2$~ns (orange dotted line), 
$\tau=1$~ns (dark olive
 green dot-dashed line), $\tau=0.5$~ns (light blue double-dot-dashed line). The adiabatic results are
 also shown  (black solid line).}
  \label{fig:function_fwhm_03_00_03_01}
\end{figure}
\begin{figure}[t]
  \centering
  \includegraphics[width=.32\textwidth]{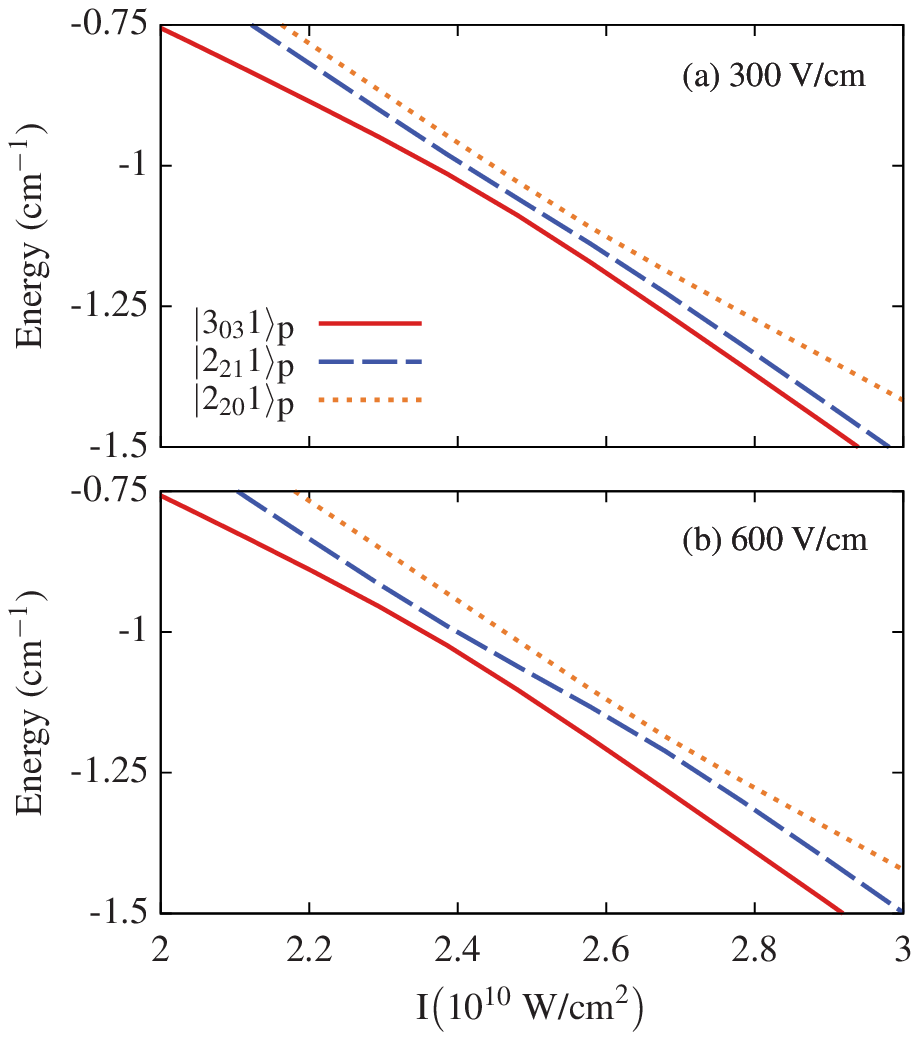}
  \caption{We show the adiabatic level structure when the pendular doublet between the 
states \ppstate{3}{0}{3}{1} and  \ppstate{2}{2}{1}{1} is formed for 
(a) $\Estatabs=\SI{300}{\fieldstrength}$ and 
(b) $\Estatabs=\SI{600}{\fieldstrength}$.}
  \label{fig:st_03_00_03_01_detail_crossing}
\end{figure}
\begin{figure}[t]
  \centering 
\includegraphics[width=.4\textwidth,angle=0]{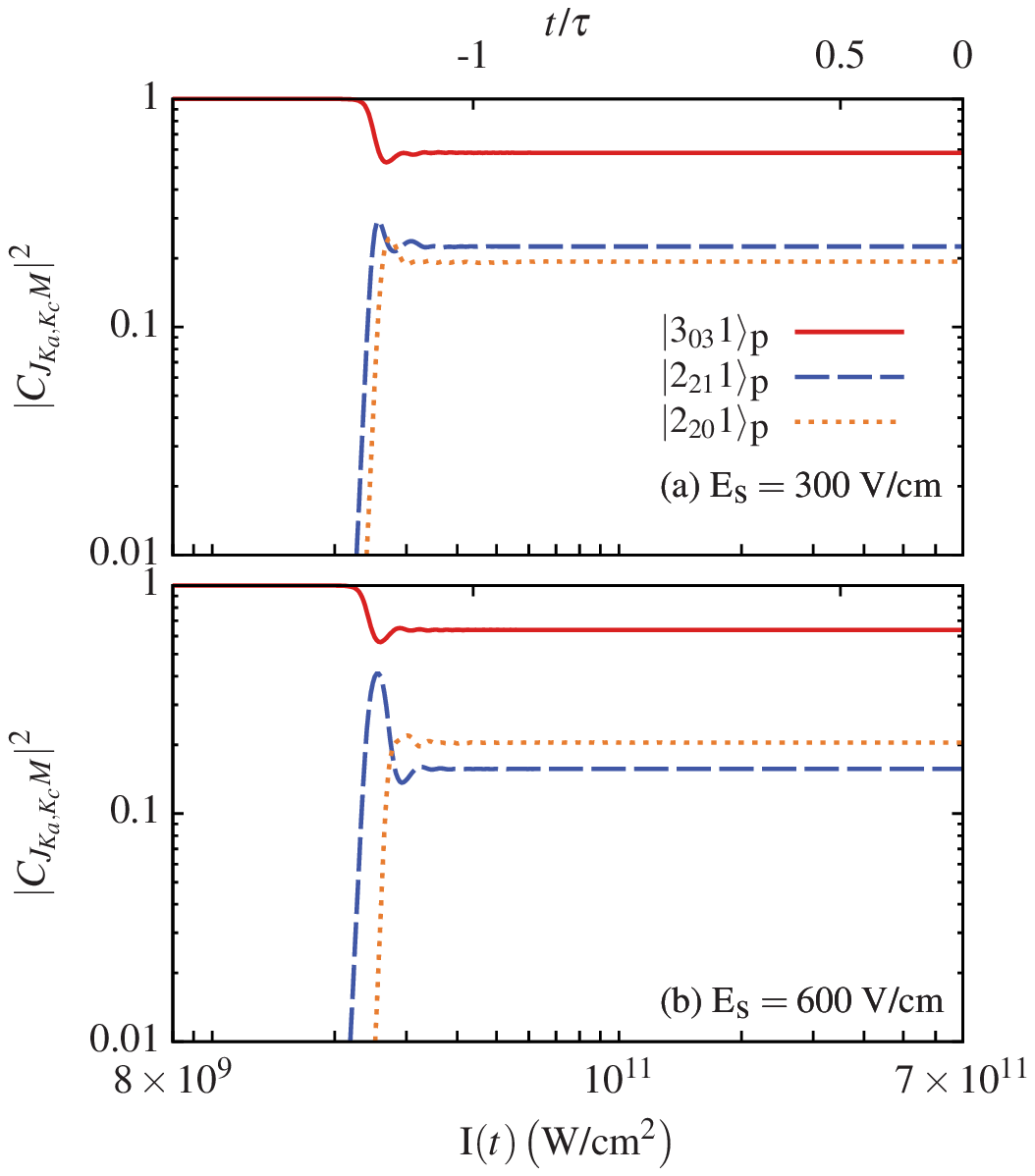}
    \caption{(Color online) 
   For  the state \pstate{3}{0}{3}{1}$_t$, 
we show the squares of the projections of the time dependent
wave function onto several adiabatic states 
versus  the laser intensity $\textup{I}(t)$ for
 dc field strengths (a) $\Estatabs=\SI{300}{\fieldstrength}$ and 
 (b) $\Estatabs=\SI{600}{\fieldstrength}$.
The Gaussian pulse has $\Ialign=\SI{7e11}{\intensity}$ and $\tau=10$~ns.
}
  \label{fig:population_10_fwhm_03_0_3_01}
\end{figure}

As a second example, we have chosen the state \pstate{3}{0}{3}{1}$_t$, which
in the field-free case is the third one with $M=1$
and even parity under $\pi$ rotation around the fields. 
% and even symmetry in the reflection of the $XZ$-plane. 
 The orientation cosine of \pstate{3}{0}{3}{1}$_t$ is presented in 
\autoref{fig:function_fwhm_03_00_03_01} for several Gaussian pulses. 
For $\Estatabs=\SI{300}{\fieldstrength}$, doubling the FWHM from $5$~ns to $10$~ns does not
provoke an enhancement on the orientation. Analogously, using 
 $\Estatabs=\SI{600}{\fieldstrength}$, the pulses with 
$\tau=2$~ns, $5$~ns and
$10$~ns give rise a similar orientation at $t=0$. 
The dynamics of  \pstate{3}{0}{3}{1}$_t$ is strongly affected by the adiabatic states
\ppstate{2}{2}{1}{1} and \ppstate{2}{2}{0}{1}. 
The adiabatic state \ppstate{3}{0}{3}{1} undergoes an avoided crossing with  \ppstate{2}{2}{1}{1}, 
see ~\autoref{fig:st_03_00_03_01_detail_crossing}, 
just before they form the second pendular pair in this irreducible representation.
The field-free states  \pstate{2}{2}{1}{1} and  \pstate{2}{2}{0}{1}
are quasi-degenerate in energy, and  in the presence of the fields, their energies 
remain very close  as $\textup{I}(t)$ is varied if the electric field is weak.
For $\Estatabs=\SI{300}{\fieldstrength}$, when the pendular doublet between \ppstate{3}{0}{3}{1}
and \ppstate{2}{2}{1}{1} is formed, 
\ppstate{2}{2}{0}{1} is energetically very close and the levels \ppstate{2}{2}{1}{1} and 
\ppstate{2}{2}{0}{1} suffer an avoided crossing, see~\autoref{fig:st_03_00_03_01_detail_crossing}. 
The population redistribution is illustrated 
in~\autoref{fig:population_10_fwhm_03_0_3_01} for the Gaussian pulse with $\tau=10$~ns and the two
dc field strengths. 
Due to the avoided crossing between \ppstate{3}{0}{3}{1} and \ppstate{2}{2}{1}{1}, 
$|C_{2_{22}1}(t)|^2$ achieves a first maximum as a function
of $\textup{I}(t)$, and afterwards it reaches a constant value once the pendular doublet is formed. 
As $\textup{I}(t)$ is increased, the second avoided crossing between \ppstate{2}{2}{1}{1} and \ppstate{2}{2}{0}{1}
is encountered, and  the adiabatic level \ppstate{2}{2}{0}{1} 
acquires  a similar population as \ppstate{2}{2}{1}{1}, and 
they get their population almost simultaneously for $\Estatabs=\SI{300}{\fieldstrength}$. 
The dynamics  of \pstate{3}{0}{3}{1}$_t$ is dominated by the adiabatic states 
\ppstate{3}{0}{3}{1}, \ppstate{2}{2}{1}{1} and \ppstate{2}{2}{0}{1}. Since
the last two states have similar population and they are oriented in opposite directions,
the final orientation at $t=0$ of \pstate{3}{0}{3}{1}$_0$ is significantly smaller than the
adiabatic prediction. The oscillations of $\costheta$ are due to the couplings between these three
pendular states.

\begin{figure}[t]
  \centering 
\includegraphics[width=.4\textwidth,angle=0]{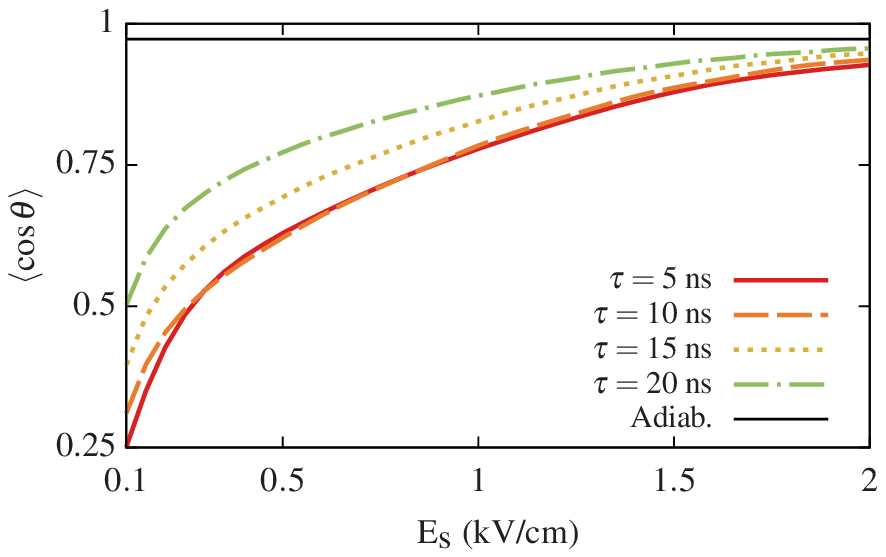}
    \caption{(Color online) 
   For  the state \pstate{3}{0}{3}{1}$_0$, 
we show the orientation at $t=0$ versus the electric field strength $\Estatabs$, for
a laser pulse with $\Ialign=\SI{7e11}{\intensity}$ and several FWHM.
}
  \label{fig:state_3031_versus_es}
\end{figure}

Finally, we investigate the rotational dynamics of this state \pstate{3}{0}{3}{1}$_t$ as the 
dc field strength is increased for several FWHM and  $\Ialign=\SI{7e11}{\intensity}$.
The orientation cosine is presented in \autoref{fig:state_3031_versus_es}. 
For the four Gaussian pulses, $\costheta$ monotonically increases as  $\Estatabs$ is enhanced.
However, even the field parameters $\tau=20$~ns and $\Estatabs=\SI{2}{\kfieldstrength}$
do not give rise to a fully
adiabatic dynamics; our time-dependent calculations provide $\costheta=0.956$, which is smaller than the
adiabatic limit $\costheta=0.973$. Note that the adiabatic value is independent of  $\Estatabs$. 
An important feature of this state is that using weak dc fields, the dynamics is not adiabatic even
if the FWHM is increased up to $20$~ns.
This lack of adiabaticity is again explained in terms of the rotational dynamics through the avoided 
crossings. 
The avoided crossing involving the adiabatic states \ppstate{3}{0}{3}{1} and \ppstate{2}{2}{1}{1}  
is crossed diabatically even for $\tau=20$~ns and 
$\Estatabs\le\SI{2}{\kfieldstrength}$. 
The second avoided crossing among \ppstate{2}{2}{1}{1} and  \ppstate{2}{2}{0}{1} 
is again passed non-adiabatically for these field configurations, and \ppstate{2}{2}{0}{1} acquires population. 
By increasing the dc field strength, the population of the state \ppstate{2}{2}{0}{1} at
$t=0$ is also increased. 
For sufficiently strong dc field,   the dynamics  of \pstate{3}{0}{3}{1}$_t$ is dominated by the adiabatic states 
\ppstate{3}{0}{3}{1} and \ppstate{2}{2}{0}{1}, and the population of  \ppstate{2}{2}{1}{1}  is reduced.
Since \ppstate{3}{0}{3}{1} and \ppstate{2}{2}{0}{1} are right-way oriented in the pendular regime, 
 $\costheta$ shows a smooth increasing behaviour as a function of $\Estatabs$.
For this state, the adiabatic dynamics is reached only if long enough Gaussian pulses
are used, and increasing the  dc field strengths will facilitate to reach this adiabatic limit. 

\section{Conclusions}
\label{sec:conclusions}

In this work, we have investigated the impact of
parallel dc fields and non-resonant laser pulses on an asymmetric top molecule.
For several rotational levels, we have explored in detail their rotational dynamics as the 
intensity of the laser pulse is increased till its peak value.  
Such a study has allowed us to identify the sources of non-adiabatic effects and the regime
when they appear.
In addition, we have analyzed the degree of orientation as the FWHM of the Gaussian
pulse and the electric field strength are varied.

We have encountered only a few rotational states, such as \pstate{0}{0}{0}{0}$_t$ and
\pstate{1}{0}{1}{1}$_t$,  for which the field-dressed dynamics is dominated by 
the formation of the pendular pairs. 
For other states, the dynamics is more complicated 
because the time evolution of their wave function is strongly affected by
the avoided crossings. 
At the ac field regime where the pendular doublets are formed, the presence of additional avoided
crossings provokes the interaction between three or even more adiabatic levels.
In such a region, the complexity of these avoided crossings prevent us from using the Landau-Zener criteria or a two state model to analyze the rotational dynamics
through them.  
The avoided crossings give rise to a highly non-adiabatic dynamics, and the final degree
of orientation could be reduced. We have shown that dc fields with strengths up to $\SI{2}{\kfieldstrength}$
do not ensure an adiabatic dynamics for low-lying rotational states.
Thus, the path to the adiabaticity necessitates laser pulses with longer temporal widths. 
Due to the different time scales associated to both phenomena, 
the pendular doublet formation and avoided crossings, the field configuration 
required to achieve an adiabatic dynamics simultaneously for many rotational states 
becomes harder to produce experimentally.
A laser pulse could be designed with a small slope of the intensity in the field regime where most of the
avoided crossings appear and pendular doublet are formed, trying to minimize the population
redistribution at that region. So that, the degree of adiabaticity could be significantly enhanced.

In this work, we have analyzed the field-dressed dynamics of benzonitrile, 
but the above-observed physical phenomena are 
expected to occur in other polar asymmetric top molecules.
Due to the complex structure of these systems,  
a similar theoretical study should be performed for each specific molecule.
The field regime under which the adiabatic dynamics would be 
achieved strongly depends on the rotational constant, the polarizability tensor and the 
permanent dipole moment.

A natural extension to this work would be the investigation  of 
an asymmetric top molecule in a nonparallel field configuration,  as 
those used in the mixed-field orientation experiments~\cite{kupper:prl102,kupper:jcp131}.
For tilted fields, the complexity of the field-dressed level structure is enhanced due to the 
presence of avoided  crossings between states having different field-free magnetic quantum 
numbers.  Due to this new source of non-adiabatic effects,
the degree of orientation in non-parallel dc and ac fields could be reduced.
In this work, we have  shown  that for parallel fields, 
the avoided crossing among states with the same
symmetry  are passed, in general, diabatically for many
field configurations. 
Thus, a time-dependent description will allow us to
revise our prediction that in titled fields 
the avoided crossings among states with the same (different) field-free value of  $M$ are  crossed
adiabatically (diabatically)~\cite{omiste:pccp2011}.

\begin{acknowledgments}
We would like to thank Jochen K\"upper and Henrik Stapelfeldt  for fruitful discussions, and
Hans-Dieter Meyer for providing us the code of the short iterative Lanczos algorithm. 
Financial support by the Spanish project FIS2011-24540 (MICINN), the
Grants P11-FQM-7276  and FQM-4643 (Junta de Andaluc\'{\i}a), and Andalusian 
research group FQM-207 is
gratefully appreciated.
This research was  supported in part by the National Science Foundation under 
Grant No. PHY11-25915.
 J.J.O. acknowledges the support of ME under the program FPU. 

\end{acknowledgments}
\bibliographystyle{apsrev}
%\bibliography{asymmetric_time}

\begin{thebibliography}{43}
\expandafter\ifx\csname natexlab\endcsname\relax\def\natexlab#1{#1}\fi
\expandafter\ifx\csname bibnamefont\endcsname\relax
  \def\bibnamefont#1{#1}\fi
\expandafter\ifx\csname bibfnamefont\endcsname\relax
  \def\bibfnamefont#1{#1}\fi
\expandafter\ifx\csname citenamefont\endcsname\relax
  \def\citenamefont#1{#1}\fi
\expandafter\ifx\csname url\endcsname\relax
  \def\url#1{\texttt{#1}}\fi
\expandafter\ifx\csname urlprefix\endcsname\relax\def\urlprefix{URL }\fi
\providecommand{\bibinfo}[2]{#2}
\providecommand{\eprint}[2][]{\url{#2}}

\bibitem[{\citenamefont{Brooks}(1976)}]{brooks:science}
\bibinfo{author}{\bibfnamefont{P.~R.} \bibnamefont{Brooks}},
  \bibinfo{journal}{Science} \textbf{\bibinfo{volume}{193}},
  \bibinfo{pages}{11} (\bibinfo{year}{1976}).

\bibitem[{\citenamefont{Brooks and Jones}(1966)}]{brooks:jcp45}
\bibinfo{author}{\bibfnamefont{P.~R.} \bibnamefont{Brooks}} \bibnamefont{and}
  \bibinfo{author}{\bibfnamefont{E.~M.} \bibnamefont{Jones}},
  \bibinfo{journal}{J. Chem. Phys.} \textbf{\bibinfo{volume}{45}},
  \bibinfo{pages}{3449} (\bibinfo{year}{1966}).

\bibitem[{\citenamefont{Loesch and M\"oller}(1992)}]{loesch:9016}
\bibinfo{author}{\bibfnamefont{H.~J.} \bibnamefont{Loesch}} \bibnamefont{and}
  \bibinfo{author}{\bibfnamefont{J.}~\bibnamefont{M\"oller}},
  \bibinfo{journal}{J. Chem. Phys.} \textbf{\bibinfo{volume}{97}},
  \bibinfo{pages}{9016} (\bibinfo{year}{1992}).

\bibitem[{\citenamefont{Aoiz et~al.}(1998)\citenamefont{Aoiz, Friedrich,
  Herrero, R\'abanos, and Verdasco}}]{aoiz:chem_phys_lett_289}
\bibinfo{author}{\bibfnamefont{F.~J.} \bibnamefont{Aoiz}},
  \bibinfo{author}{\bibfnamefont{B.}~\bibnamefont{Friedrich}},
  \bibinfo{author}{\bibfnamefont{V.~J.} \bibnamefont{Herrero}},
  \bibinfo{author}{\bibfnamefont{V.~S.} \bibnamefont{R\'abanos}},
  \bibnamefont{and} \bibinfo{author}{\bibfnamefont{J.~E.}
  \bibnamefont{Verdasco}}, \bibinfo{journal}{Chem. Phys. Lett.}
  \textbf{\bibinfo{volume}{289}}, \bibinfo{pages}{132 } (\bibinfo{year}{1998}).

\bibitem[{\citenamefont{Aquilanti et~al.}(2005)\citenamefont{Aquilanti,
  Bartolomei, Pirani, Cappelletti, and Vecchiocattivi}}]{aquilanti:pccp_7}
\bibinfo{author}{\bibfnamefont{V.}~\bibnamefont{Aquilanti}},
  \bibinfo{author}{\bibfnamefont{M.}~\bibnamefont{Bartolomei}},
  \bibinfo{author}{\bibfnamefont{F.}~\bibnamefont{Pirani}},
  \bibinfo{author}{\bibfnamefont{D.}~\bibnamefont{Cappelletti}},
  \bibnamefont{and}
  \bibinfo{author}{\bibfnamefont{F.}~\bibnamefont{Vecchiocattivi}},
  \bibinfo{journal}{Phys. Chem. Chem. Phys} \textbf{\bibinfo{volume}{7}},
  \bibinfo{pages}{291} (\bibinfo{year}{2005}).

\bibitem[{\citenamefont{Bisgaard et~al.}(2009)\citenamefont{Bisgaard, Clarkin,
  Wu, Lee, Gessner, Hayden, and Stolow}}]{Bisgaard:Science323:1464}
\bibinfo{author}{\bibfnamefont{C.~Z.} \bibnamefont{Bisgaard}},
  \bibinfo{author}{\bibfnamefont{O.~J.} \bibnamefont{Clarkin}},
  \bibinfo{author}{\bibfnamefont{G.}~\bibnamefont{Wu}},
  \bibinfo{author}{\bibfnamefont{A.~M.~D.} \bibnamefont{Lee}},
  \bibinfo{author}{\bibfnamefont{O.}~\bibnamefont{Gessner}},
  \bibinfo{author}{\bibfnamefont{C.~C.} \bibnamefont{Hayden}},
  \bibnamefont{and} \bibinfo{author}{\bibfnamefont{A.}~\bibnamefont{Stolow}},
  \bibinfo{journal}{Science} \textbf{\bibinfo{volume}{323}},
  \bibinfo{pages}{1464} (\bibinfo{year}{2009}).

\bibitem[{\citenamefont{Holmegaard et~al.}(2010)\citenamefont{Holmegaard,
  Hansen, Kalhoj, Kragh, Stapelfeldt, Filsinger, K\"upper, Meijer, Dimitrovski,
  Abu-samha et~al.}}]{Holmegaard:natphys6}
\bibinfo{author}{\bibfnamefont{L.}~\bibnamefont{Holmegaard}},
  \bibinfo{author}{\bibfnamefont{J.~L.} \bibnamefont{Hansen}},
  \bibinfo{author}{\bibfnamefont{L.}~\bibnamefont{Kalhoj}},
  \bibinfo{author}{\bibfnamefont{S.~L.} \bibnamefont{Kragh}},
  \bibinfo{author}{\bibfnamefont{H.}~\bibnamefont{Stapelfeldt}},
  \bibinfo{author}{\bibfnamefont{F.}~\bibnamefont{Filsinger}},
  \bibinfo{author}{\bibfnamefont{J.}~\bibnamefont{K\"upper}},
  \bibinfo{author}{\bibfnamefont{G.}~\bibnamefont{Meijer}},
  \bibinfo{author}{\bibfnamefont{D.}~\bibnamefont{Dimitrovski}},
  \bibinfo{author}{\bibfnamefont{M.}~\bibnamefont{Abu-samha}},
  \bibnamefont{et~al.}, \bibinfo{journal}{Nat. Phys.}
  \textbf{\bibinfo{volume}{6}}, \bibinfo{pages}{428} (\bibinfo{year}{2010}).

\bibitem[{\citenamefont{Hansen et~al.}(2011{\natexlab{a}})\citenamefont{Hansen,
  Stapelfeldt, Dimitrovski, Abu-samha, Martiny, and
  Madsen}}]{Hansen:PRL106:073001}
\bibinfo{author}{\bibfnamefont{J.~L.} \bibnamefont{Hansen}},
  \bibinfo{author}{\bibfnamefont{H.}~\bibnamefont{Stapelfeldt}},
  \bibinfo{author}{\bibfnamefont{D.}~\bibnamefont{Dimitrovski}},
  \bibinfo{author}{\bibfnamefont{M.}~\bibnamefont{Abu-samha}},
  \bibinfo{author}{\bibfnamefont{C.~P.~J.} \bibnamefont{Martiny}},
  \bibnamefont{and} \bibinfo{author}{\bibfnamefont{L.~B.}
  \bibnamefont{Madsen}}, \bibinfo{journal}{Phys. Rev. Lett.}
  \textbf{\bibinfo{volume}{106}}, \bibinfo{pages}{073001}
  (\bibinfo{year}{2011}{\natexlab{a}}).

\bibitem[{\citenamefont{Frumker et~al.}(2012)\citenamefont{Frumker, Hebeisen,
  Kajumba, Bertrand, W\"orner, Spanner, Villeneuve, Naumov, and
  Corkum}}]{frumker2012}
\bibinfo{author}{\bibfnamefont{E.}~\bibnamefont{Frumker}},
  \bibinfo{author}{\bibfnamefont{C.~T.} \bibnamefont{Hebeisen}},
  \bibinfo{author}{\bibfnamefont{N.}~\bibnamefont{Kajumba}},
  \bibinfo{author}{\bibfnamefont{J.~B.} \bibnamefont{Bertrand}},
  \bibinfo{author}{\bibfnamefont{H.~J.} \bibnamefont{W\"orner}},
  \bibinfo{author}{\bibfnamefont{M.}~\bibnamefont{Spanner}},
  \bibinfo{author}{\bibfnamefont{D.~M.} \bibnamefont{Villeneuve}},
  \bibinfo{author}{\bibfnamefont{A.}~\bibnamefont{Naumov}}, \bibnamefont{and}
  \bibinfo{author}{\bibfnamefont{P.~B.} \bibnamefont{Corkum}},
  \bibinfo{journal}{Phys. Rev. Lett.} \textbf{\bibinfo{volume}{109}},
  \bibinfo{pages}{113901} (\bibinfo{year}{2012}).

\bibitem[{\citenamefont{Kraus et~al.}(2012)\citenamefont{Kraus, Rupenyan, and
  W\"orner}}]{kraus2012}
\bibinfo{author}{\bibfnamefont{P.~M.} \bibnamefont{Kraus}},
  \bibinfo{author}{\bibfnamefont{A.}~\bibnamefont{Rupenyan}}, \bibnamefont{and}
  \bibinfo{author}{\bibfnamefont{H.~J.} \bibnamefont{W\"orner}},
  \bibinfo{journal}{Phys. Rev. Lett.} \textbf{\bibinfo{volume}{109}},
  \bibinfo{pages}{233903} (\bibinfo{year}{2012}).

\bibitem[{\citenamefont{Spector et~al.}(2012)\citenamefont{Spector, Artamonov,
  Miyabe, Martinez, Seideman, Guehr, and Bucksbaum}}]{buckbaum}
\bibinfo{author}{\bibfnamefont{L.~S.} \bibnamefont{Spector}},
  \bibinfo{author}{\bibfnamefont{M.}~\bibnamefont{Artamonov}},
  \bibinfo{author}{\bibfnamefont{S.}~\bibnamefont{Miyabe}},
  \bibinfo{author}{\bibfnamefont{T.}~\bibnamefont{Martinez}},
  \bibinfo{author}{\bibfnamefont{T.}~\bibnamefont{Seideman}},
  \bibinfo{author}{\bibfnamefont{M.}~\bibnamefont{Guehr}}, \bibnamefont{and}
  \bibinfo{author}{\bibfnamefont{P.~H.} \bibnamefont{Bucksbaum}}
  (\bibinfo{year}{2012}), \bibinfo{note}{arXiv:1207.2517}.

\bibitem[{\citenamefont{Stolte}(1988)}]{stolte}
\bibinfo{author}{\bibfnamefont{S.}~\bibnamefont{Stolte}},
  \emph{\bibinfo{title}{Atomic and Molecular Beam Methods}}
  (\bibinfo{publisher}{Oxford University Press, New York},
  \bibinfo{year}{1988}).

\bibitem[{\citenamefont{Loesch and Remscheid}(1990)}]{loesch:jcp93}
\bibinfo{author}{\bibfnamefont{H.~J.} \bibnamefont{Loesch}} \bibnamefont{and}
  \bibinfo{author}{\bibfnamefont{A.}~\bibnamefont{Remscheid}},
  \bibinfo{journal}{J. Chem. Phys.} \textbf{\bibinfo{volume}{93}},
  \bibinfo{pages}{4779} (\bibinfo{year}{1990}).

\bibitem[{\citenamefont{Friedrich and Herschbach}(1991)}]{friedrich:nature353}
\bibinfo{author}{\bibfnamefont{B.}~\bibnamefont{Friedrich}} \bibnamefont{and}
  \bibinfo{author}{\bibfnamefont{D.~R.} \bibnamefont{Herschbach}},
  \bibinfo{journal}{Nature} \textbf{\bibinfo{volume}{353}},
  \bibinfo{pages}{412} (\bibinfo{year}{1991}).

\bibitem[{\citenamefont{Friedrich et~al.}(1991)\citenamefont{Friedrich,
  Pullman, and Herschbach}}]{friedrich:jpc95}
\bibinfo{author}{\bibfnamefont{B.}~\bibnamefont{Friedrich}},
  \bibinfo{author}{\bibfnamefont{D.~P.} \bibnamefont{Pullman}},
  \bibnamefont{and} \bibinfo{author}{\bibfnamefont{D.~R.}
  \bibnamefont{Herschbach}}, \bibinfo{journal}{J. Phys. Chem.}
  \textbf{\bibinfo{volume}{95}}, \bibinfo{pages}{8118} (\bibinfo{year}{1991}).

\bibitem[{\citenamefont{Block et~al.}(1992)\citenamefont{Block, Bohac, and
  Miller}}]{block:prl68}
\bibinfo{author}{\bibfnamefont{P.~A.} \bibnamefont{Block}},
  \bibinfo{author}{\bibfnamefont{E.~J.} \bibnamefont{Bohac}}, \bibnamefont{and}
  \bibinfo{author}{\bibfnamefont{R.~E.} \bibnamefont{Miller}},
  \bibinfo{journal}{Phys. Rev. Lett.} \textbf{\bibinfo{volume}{68}},
  \bibinfo{pages}{1303} (\bibinfo{year}{1992}).

\bibitem[{\citenamefont{Friedrich et~al.}(1992)\citenamefont{Friedrich, Rubahn,
  and Sathyamurthy}}]{friedrich:prl69}
\bibinfo{author}{\bibfnamefont{B.}~\bibnamefont{Friedrich}},
  \bibinfo{author}{\bibfnamefont{H.~G.} \bibnamefont{Rubahn}},
  \bibnamefont{and}
  \bibinfo{author}{\bibfnamefont{N.}~\bibnamefont{Sathyamurthy}},
  \bibinfo{journal}{Phys. Rev. Lett.} \textbf{\bibinfo{volume}{69}},
  \bibinfo{pages}{2487} (\bibinfo{year}{1992}).

\bibitem[{\citenamefont{Slenczka et~al.}(1994)\citenamefont{Slenczka,
  Friedrich, and Herschbach}}]{slenczka:prl72}
\bibinfo{author}{\bibfnamefont{A.}~\bibnamefont{Slenczka}},
  \bibinfo{author}{\bibfnamefont{B.}~\bibnamefont{Friedrich}},
  \bibnamefont{and}
  \bibinfo{author}{\bibfnamefont{D.}~\bibnamefont{Herschbach}},
  \bibinfo{journal}{Phys. Rev. Lett.} \textbf{\bibinfo{volume}{72}},
  \bibinfo{pages}{1806} (\bibinfo{year}{1994}).

\bibitem[{\citenamefont{Friedrich and
  Herschbach}(1999{\natexlab{a}})}]{friedrich:jcp111}
\bibinfo{author}{\bibfnamefont{B.}~\bibnamefont{Friedrich}} \bibnamefont{and}
  \bibinfo{author}{\bibfnamefont{D.~R.} \bibnamefont{Herschbach}},
  \bibinfo{journal}{J. Chem. Phys.} \textbf{\bibinfo{volume}{111}},
  \bibinfo{pages}{6157} (\bibinfo{year}{1999}{\natexlab{a}}).

\bibitem[{\citenamefont{Friedrich and
  Herschbach}(1999{\natexlab{b}})}]{friedrich:jpca103-a}
\bibinfo{author}{\bibfnamefont{B.}~\bibnamefont{Friedrich}} \bibnamefont{and}
  \bibinfo{author}{\bibfnamefont{D.}~\bibnamefont{Herschbach}},
  \bibinfo{journal}{J. Phys. Chem. A} \textbf{\bibinfo{volume}{103}},
  \bibinfo{pages}{10280} (\bibinfo{year}{1999}{\natexlab{b}}).

\bibitem[{\citenamefont{Holmegaard et~al.}(2009)\citenamefont{Holmegaard,
  Nielsen, Nevo, Stapelfeldt, Filsinger, K\"upper, and Meijer}}]{kupper:prl102}
\bibinfo{author}{\bibfnamefont{L.}~\bibnamefont{Holmegaard}},
  \bibinfo{author}{\bibfnamefont{J.~H.} \bibnamefont{Nielsen}},
  \bibinfo{author}{\bibfnamefont{I.}~\bibnamefont{Nevo}},
  \bibinfo{author}{\bibfnamefont{H.}~\bibnamefont{Stapelfeldt}},
  \bibinfo{author}{\bibfnamefont{F.}~\bibnamefont{Filsinger}},
  \bibinfo{author}{\bibfnamefont{J.}~\bibnamefont{K\"upper}}, \bibnamefont{and}
  \bibinfo{author}{\bibfnamefont{G.}~\bibnamefont{Meijer}},
  \bibinfo{journal}{Phys. Rev. Lett.} \textbf{\bibinfo{volume}{102}},
  \bibinfo{pages}{023001} (\bibinfo{year}{2009}).

\bibitem[{\citenamefont{Nevo et~al.}(2009)\citenamefont{Nevo, Holmegaard,
  Nielsen, Hansen, Stapelfeldt, Filsinger, Meijer, and K\"upper}}]{nevo:pccp11}
\bibinfo{author}{\bibfnamefont{I.}~\bibnamefont{Nevo}},
  \bibinfo{author}{\bibfnamefont{L.}~\bibnamefont{Holmegaard}},
  \bibinfo{author}{\bibfnamefont{J.}~\bibnamefont{Nielsen}},
  \bibinfo{author}{\bibfnamefont{J.~L.} \bibnamefont{Hansen}},
  \bibinfo{author}{\bibfnamefont{H.}~\bibnamefont{Stapelfeldt}},
  \bibinfo{author}{\bibfnamefont{F.}~\bibnamefont{Filsinger}},
  \bibinfo{author}{\bibfnamefont{G.}~\bibnamefont{Meijer}}, \bibnamefont{and}
  \bibinfo{author}{\bibfnamefont{J.}~\bibnamefont{K\"upper}},
  \bibinfo{journal}{Phys. Chem. Chem. Phys.} \textbf{\bibinfo{volume}{11}},
  \bibinfo{pages}{9912} (\bibinfo{year}{2009}).

\bibitem[{\citenamefont{Seideman and Hamilton}(2006)}]{seideman2006}
\bibinfo{author}{\bibfnamefont{T.}~\bibnamefont{Seideman}} \bibnamefont{and}
  \bibinfo{author}{\bibfnamefont{E.}~\bibnamefont{Hamilton}},
  \bibinfo{journal}{Adv. Atom. Mol. Opt. Phys.} \textbf{\bibinfo{volume}{52}},
  \bibinfo{pages}{289} (\bibinfo{year}{2006}).

\bibitem[{\citenamefont{Omiste et~al.}(2011{\natexlab{a}})\citenamefont{Omiste,
  G\"arttner, Schmelcher, Gonz\'{a}lez-F\'{e}rez, Holmegaard, Nielsen,
  Stapelfeldt, and K\"upper}}]{omiste:pccp2011}
\bibinfo{author}{\bibfnamefont{J.~J.} \bibnamefont{Omiste}},
  \bibinfo{author}{\bibfnamefont{M.}~\bibnamefont{G\"arttner}},
  \bibinfo{author}{\bibfnamefont{P.}~\bibnamefont{Schmelcher}},
  \bibinfo{author}{\bibfnamefont{R.}~\bibnamefont{Gonz\'{a}lez-F\'{e}rez}},
  \bibinfo{author}{\bibfnamefont{L.}~\bibnamefont{Holmegaard}},
  \bibinfo{author}{\bibfnamefont{J.~H.} \bibnamefont{Nielsen}},
  \bibinfo{author}{\bibfnamefont{H.}~\bibnamefont{Stapelfeldt}},
  \bibnamefont{and} \bibinfo{author}{\bibfnamefont{J.}~\bibnamefont{K\"upper}},
  \bibinfo{journal}{Phys. Chem. Chem. Phys.} \textbf{\bibinfo{volume}{13}},
  \bibinfo{pages}{18815} (\bibinfo{year}{2011}{\natexlab{a}}).

\bibitem[{\citenamefont{Nielsen et~al.}(2012)\citenamefont{Nielsen,
  Stapelfeldt, K\"upper, Friedrich, Omiste, and
  Gonz\'alez-F\'erez}}]{nielsen:prl2012}
\bibinfo{author}{\bibfnamefont{J.~H.} \bibnamefont{Nielsen}},
  \bibinfo{author}{\bibfnamefont{H.}~\bibnamefont{Stapelfeldt}},
  \bibinfo{author}{\bibfnamefont{J.}~\bibnamefont{K\"upper}},
  \bibinfo{author}{\bibfnamefont{B.}~\bibnamefont{Friedrich}},
  \bibinfo{author}{\bibfnamefont{J.~J.} \bibnamefont{Omiste}},
  \bibnamefont{and}
  \bibinfo{author}{\bibfnamefont{R.}~\bibnamefont{Gonz\'alez-F\'erez}},
  \bibinfo{journal}{Phys. Rev. Lett.} \textbf{\bibinfo{volume}{108}},
  \bibinfo{pages}{193001} (\bibinfo{year}{2012}).

\bibitem[{\citenamefont{Omiste and Gonz\'alez-F\'erez}(2012)}]{omiste:pra2012}
\bibinfo{author}{\bibfnamefont{J.~J.} \bibnamefont{Omiste}} \bibnamefont{and}
  \bibinfo{author}{\bibfnamefont{R.}~\bibnamefont{Gonz\'alez-F\'erez}},
  \bibinfo{journal}{Phys. Rev. A} \textbf{\bibinfo{volume}{86}},
  \bibinfo{pages}{043437} (\bibinfo{year}{2012}).

\bibitem[{\citenamefont{Filsinger et~al.}(2009)\citenamefont{Filsinger,
  K\"upper, Meijer, Holmegaard, Nielsen, Nevo, Hansen, and
  Stapelfeldt}}]{kupper:jcp131}
\bibinfo{author}{\bibfnamefont{F.}~\bibnamefont{Filsinger}},
  \bibinfo{author}{\bibfnamefont{J.}~\bibnamefont{K\"upper}},
  \bibinfo{author}{\bibfnamefont{G.}~\bibnamefont{Meijer}},
  \bibinfo{author}{\bibfnamefont{L.}~\bibnamefont{Holmegaard}},
  \bibinfo{author}{\bibfnamefont{J.~H.} \bibnamefont{Nielsen}},
  \bibinfo{author}{\bibfnamefont{I.}~\bibnamefont{Nevo}},
  \bibinfo{author}{\bibfnamefont{J.~L.} \bibnamefont{Hansen}},
  \bibnamefont{and}
  \bibinfo{author}{\bibfnamefont{H.}~\bibnamefont{Stapelfeldt}},
  \bibinfo{journal}{J. Chem. Phys.} \textbf{\bibinfo{volume}{131}},
  \bibinfo{pages}{064309} (\bibinfo{year}{2009}).

\bibitem[{\citenamefont{Bulthuis et~al.}(1997)\citenamefont{Bulthuis, Miller,
  and Loesch}}]{bulthuis:jpca101}
\bibinfo{author}{\bibfnamefont{J.}~\bibnamefont{Bulthuis}},
  \bibinfo{author}{\bibfnamefont{J.}~\bibnamefont{Miller}}, \bibnamefont{and}
  \bibinfo{author}{\bibfnamefont{H.~J.} \bibnamefont{Loesch}},
  \bibinfo{journal}{J. Phys. Chem. A} \textbf{\bibinfo{volume}{101}},
  \bibinfo{pages}{7684} (\bibinfo{year}{1997}).

\bibitem[{\citenamefont{Escribano et~al.}(2000)\citenamefont{Escribano, Mat\'e,
  Ortigoso, and Ortigoso}}]{escribano:pra62}
\bibinfo{author}{\bibfnamefont{R.}~\bibnamefont{Escribano}},
  \bibinfo{author}{\bibfnamefont{B.}~\bibnamefont{Mat\'e}},
  \bibinfo{author}{\bibfnamefont{F.}~\bibnamefont{Ortigoso}}, \bibnamefont{and}
  \bibinfo{author}{\bibfnamefont{J.}~\bibnamefont{Ortigoso}},
  \bibinfo{journal}{Phys. Rev. A} \textbf{\bibinfo{volume}{62}},
  \bibinfo{pages}{023407} (\bibinfo{year}{2000}).

\bibitem[{\citenamefont{Schwettman et~al.}(2005)\citenamefont{Schwettman,
  Franklin, Overstreet, and Shaffer}}]{schwettman:jcp123}
\bibinfo{author}{\bibfnamefont{A.}~\bibnamefont{Schwettman}},
  \bibinfo{author}{\bibfnamefont{J.}~\bibnamefont{Franklin}},
  \bibinfo{author}{\bibfnamefont{K.~R.} \bibnamefont{Overstreet}},
  \bibnamefont{and} \bibinfo{author}{\bibfnamefont{J.~P.}
  \bibnamefont{Shaffer}}, \bibinfo{journal}{J. Chem. Phys.}
  \textbf{\bibinfo{volume}{123}}, \bibinfo{pages}{194305}
  (\bibinfo{year}{2005}).

\bibitem[{\citenamefont{Wohlfart et~al.}(2008)\citenamefont{Wohlfart,
  Filsinger, Gr\"atz, K\"upper, and Meijer}}]{wohlfart:phys_rev_a_78}
\bibinfo{author}{\bibfnamefont{K.}~\bibnamefont{Wohlfart}},
  \bibinfo{author}{\bibfnamefont{F.}~\bibnamefont{Filsinger}},
  \bibinfo{author}{\bibfnamefont{F.}~\bibnamefont{Gr\"atz}},
  \bibinfo{author}{\bibfnamefont{J.}~\bibnamefont{K\"upper}}, \bibnamefont{and}
  \bibinfo{author}{\bibfnamefont{G.}~\bibnamefont{Meijer}},
  \bibinfo{journal}{Phys. Rev. A} \textbf{\bibinfo{volume}{78}},
  \bibinfo{pages}{033421} (\bibinfo{year}{2008}).

\bibitem[{\citenamefont{Kirste et~al.}(2009)\citenamefont{Kirste, Sartakov,
  Schnell, and Meijer}}]{kirste:phys_rev_a_79}
\bibinfo{author}{\bibfnamefont{M.}~\bibnamefont{Kirste}},
  \bibinfo{author}{\bibfnamefont{B.~G.} \bibnamefont{Sartakov}},
  \bibinfo{author}{\bibfnamefont{M.}~\bibnamefont{Schnell}}, \bibnamefont{and}
  \bibinfo{author}{\bibfnamefont{G.}~\bibnamefont{Meijer}},
  \bibinfo{journal}{Phys. Rev. A} \textbf{\bibinfo{volume}{79}},
  \bibinfo{pages}{051401} (\bibinfo{year}{2009}).

\bibitem[{\citenamefont{Meek et~al.}(2009)\citenamefont{Meek, Conrad, and
  Meijer}}]{Meek2009}
\bibinfo{author}{\bibfnamefont{S.~A.} \bibnamefont{Meek}},
  \bibinfo{author}{\bibfnamefont{H.}~\bibnamefont{Conrad}}, \bibnamefont{and}
  \bibinfo{author}{\bibfnamefont{G.}~\bibnamefont{Meijer}},
  \bibinfo{journal}{Science} \textbf{\bibinfo{volume}{324}},
  \bibinfo{pages}{1699} (\bibinfo{year}{2009}).

\bibitem[{\citenamefont{Wall et~al.}(2010)\citenamefont{Wall, Tokunaga, Hinds,
  and Tarbutt}}]{wall:phys_rev_a_81}
\bibinfo{author}{\bibfnamefont{T.~E.} \bibnamefont{Wall}},
  \bibinfo{author}{\bibfnamefont{S.~K.} \bibnamefont{Tokunaga}},
  \bibinfo{author}{\bibfnamefont{E.~A.} \bibnamefont{Hinds}}, \bibnamefont{and}
  \bibinfo{author}{\bibfnamefont{M.~R.} \bibnamefont{Tarbutt}},
  \bibinfo{journal}{Phys. Rev. A} \textbf{\bibinfo{volume}{81}},
  \bibinfo{pages}{033414} (\bibinfo{year}{2010}).

\bibitem[{\citenamefont{Zare}(1988)}]{zare}
\bibinfo{author}{\bibfnamefont{R.~N.} \bibnamefont{Zare}},
  \emph{\bibinfo{title}{Angular {M}omentum: {U}nderstanding {S}patial {A}spects
  in {C}hemistry and {P}hysics}} (\bibinfo{publisher}{John Wiley and Sons, New
  York}, \bibinfo{year}{1988}).

\bibitem[{\citenamefont{King et~al.}(1943)\citenamefont{King, Hainer, and
  Cross}}]{king_jcp11}
\bibinfo{author}{\bibfnamefont{G.~W.} \bibnamefont{King}},
  \bibinfo{author}{\bibfnamefont{R.~M.} \bibnamefont{Hainer}},
  \bibnamefont{and} \bibinfo{author}{\bibfnamefont{P.~C.} \bibnamefont{Cross}},
  \bibinfo{journal}{J. Chem. Phys.} \textbf{\bibinfo{volume}{11}},
  \bibinfo{pages}{27} (\bibinfo{year}{1943}).

\bibitem[{\citenamefont{Beck et~al.}(2000)\citenamefont{Beck, J\"ackle, Worth,
  and Meyer}}]{Beck:phys_rep_324}
\bibinfo{author}{\bibfnamefont{M.}~\bibnamefont{Beck}},
  \bibinfo{author}{\bibfnamefont{A.}~\bibnamefont{J\"ackle}},
  \bibinfo{author}{\bibfnamefont{G.}~\bibnamefont{Worth}}, \bibnamefont{and}
  \bibinfo{author}{\bibfnamefont{H.-D.} \bibnamefont{Meyer}},
  \bibinfo{journal}{Physics Reports} \textbf{\bibinfo{volume}{324}},
  \bibinfo{pages}{1 } (\bibinfo{year}{2000}).

\bibitem[{\citenamefont{Omiste et~al.}(2011{\natexlab{b}})\citenamefont{Omiste,
  Gonz\'alez-F\'erez, and Schmelcher}}]{omiste:jcp2011}
\bibinfo{author}{\bibfnamefont{J.~J.} \bibnamefont{Omiste}},
  \bibinfo{author}{\bibfnamefont{R.}~\bibnamefont{Gonz\'alez-F\'erez}},
  \bibnamefont{and}
  \bibinfo{author}{\bibfnamefont{P.}~\bibnamefont{Schmelcher}},
  \bibinfo{journal}{J. Chem. Phys.} \textbf{\bibinfo{volume}{135}},
  \bibinfo{pages}{064310} (\bibinfo{year}{2011}{\natexlab{b}}).

\bibitem[{\citenamefont{Ballentine}(1998)}]{ballentine:quantum_mechanics}
\bibinfo{author}{\bibfnamefont{L.~B.} \bibnamefont{Ballentine}},
  \emph{\bibinfo{title}{Quantum {M}echanics: {A} {M}odern {D}evelopment}}
  (\bibinfo{publisher}{World Scientific, Singapore}, \bibinfo{year}{1998}).

\bibitem[{\citenamefont{Wohlfart et~al.}(2007)\citenamefont{Wohlfart, Schnell,
  Grabow, and K\"upper}}]{wohlfart:jms247}
\bibinfo{author}{\bibfnamefont{K.}~\bibnamefont{Wohlfart}},
  \bibinfo{author}{\bibfnamefont{M.}~\bibnamefont{Schnell}},
  \bibinfo{author}{\bibfnamefont{J.-U.} \bibnamefont{Grabow}},
  \bibnamefont{and} \bibinfo{author}{\bibfnamefont{J.}~\bibnamefont{K\"upper}},
  \bibinfo{journal}{J. Mol. Spectrosc} \textbf{\bibinfo{volume}{247}},
  \bibinfo{pages}{119} (\bibinfo{year}{2007}).

\bibitem[{\citenamefont{Hansen et~al.}(2011{\natexlab{b}})\citenamefont{Hansen,
  Holmegaard, Kalh\o{}j, Kragh, Stapelfeldt, Filsinger, Meijer, K\"upper,
  Dimitrovski, Abu-samha et~al.}}]{PhysRevA.83.023406}
\bibinfo{author}{\bibfnamefont{J.~L.} \bibnamefont{Hansen}},
  \bibinfo{author}{\bibfnamefont{L.}~\bibnamefont{Holmegaard}},
  \bibinfo{author}{\bibfnamefont{L.}~\bibnamefont{Kalh\o{}j}},
  \bibinfo{author}{\bibfnamefont{S.~L.} \bibnamefont{Kragh}},
  \bibinfo{author}{\bibfnamefont{H.}~\bibnamefont{Stapelfeldt}},
  \bibinfo{author}{\bibfnamefont{F.}~\bibnamefont{Filsinger}},
  \bibinfo{author}{\bibfnamefont{G.}~\bibnamefont{Meijer}},
  \bibinfo{author}{\bibfnamefont{J.}~\bibnamefont{K\"upper}},
  \bibinfo{author}{\bibfnamefont{D.}~\bibnamefont{Dimitrovski}},
  \bibinfo{author}{\bibfnamefont{M.}~\bibnamefont{Abu-samha}},
  \bibnamefont{et~al.}, \bibinfo{journal}{Phys. Rev. A}
  \textbf{\bibinfo{volume}{83}}, \bibinfo{pages}{023406}
  (\bibinfo{year}{2011}{\natexlab{b}}).

\bibitem[{\citenamefont{Sugawara et~al.}(2008)\citenamefont{Sugawara, Goban,
  Minemoto, and Sakai}}]{Sugawara2008}
\bibinfo{author}{\bibfnamefont{Y.}~\bibnamefont{Sugawara}},
  \bibinfo{author}{\bibfnamefont{A.}~\bibnamefont{Goban}},
  \bibinfo{author}{\bibfnamefont{S.}~\bibnamefont{Minemoto}}, \bibnamefont{and}
  \bibinfo{author}{\bibfnamefont{H.}~\bibnamefont{Sakai}},
  \bibinfo{journal}{Phys. Rev. A} \textbf{\bibinfo{volume}{77}},
  \bibinfo{pages}{031403(R)} (\bibinfo{year}{2008}).

\bibitem[{\citenamefont{Muramatsu et~al.}(2009)\citenamefont{Muramatsu, Hita,
  Minemoto, and Sakai}}]{Muramatsu2009}
\bibinfo{author}{\bibfnamefont{M.}~\bibnamefont{Muramatsu}},
  \bibinfo{author}{\bibfnamefont{M.}~\bibnamefont{Hita}},
  \bibinfo{author}{\bibfnamefont{S.}~\bibnamefont{Minemoto}}, \bibnamefont{and}
  \bibinfo{author}{\bibfnamefont{H.}~\bibnamefont{Sakai}},
  \bibinfo{journal}{Phys. Rev. A} \textbf{\bibinfo{volume}{79}},
  \bibinfo{pages}{011403(R)} (\bibinfo{year}{2009}).

\end{thebibliography}

\end{document}